\documentclass[10pt]{iopart}
\usepackage[symbol]{footmisc}

\expandafter\let\csname equation*\endcsname\relax
\expandafter\let\csname endequation*\endcsname\relax
\usepackage{amsmath}
\usepackage{amsfonts}
\usepackage{xcolor}
\usepackage{graphicx}
\usepackage{hyperref}
\usepackage{cite}
\usepackage[normalem]{ulem}
\usepackage{bm}

\DeclareMathOperator\erfc{Erfc}
\usepackage{mathtools}
\newcommand\underrel[3][]{\mathrel{\mathop{#3}\limits_{%
			\ifx c#1\relax\mathclap{#2}\else#2\fi}}}

\begin{document}
	
\title[Non-equilibrium thermodynamics of fluctuating potentials]{Non-equilibrium thermodynamics of diffusion in fluctuating potentials}
	
\author{Henry Alston\footnotemark[1]$^{1}$, Luca Cocconi\footnotemark[\value{footnote}]$^{1,2}$ and Thibault Bertrand$^{1}$}
\footnotetext{ These authors contributed equally to the work.}
\address{$^{1}$ Department of Mathematics, Imperial College London, South Kensington, London SW7 2BZ, United Kingdom}	
\address{$^{2}$ The Francis Crick Institute, London NW1 1AT, United Kingdom}
\ead{t.bertrand@imperial.ac.uk}
	
\vspace{10pt}
\begin{indented}
	\item[]
\end{indented}

\begin{abstract}
A positive rate of entropy production at steady state is a distinctive feature of truly non-equilibrium processes. Exact results, while being often limited to simple models, offer a unique opportunity to explore the thermodynamic features of these processes in full details. Here we derive analytical results for the steady-state rate of entropy production in single particle systems driven away from equilibrium by the fluctuations of an external potential of arbitrary shapes. Subsequently, we provide exact results for a diffusive particle in a harmonic trap whose potential stiffness varies in time according to both discrete and continuous Markov processes. In particular, studying the case of a fully intermittent potential allows us to introduce an effective model of stochastic resetting for which it is possible to obtain finite non-negative entropy production. Altogether, this work lays the foundation for a non-equilibrium thermodynamic theory of fluctuating potentials, with immediate applications to stochastic resetting processes, fluctuations in optical traps and fluctuating interactions in living systems.
\end{abstract}
	


 

\section{Introduction} 
\label{sec:intro}

Stochastic thermodynamics represents one of the most powerful tools at our disposal in the effort to characterize generic properties of non-equilibrium processes. It provides a framework to extend the ideas of traditional thermodynamics to regimes and scales where some of the assumptions underlying the latter theory break down \cite{Peliti2021,Seifert2012,Cocconi2020}. In particular, the possibility of developing a thermodynamically-consistent description of mesoscopic systems subject to non-negligible noise (a paradigmatic example being overdamped colloidal particles) has unveiled a wealth of fascinating relations among the fluctuating counterparts of traditional thermodynamic observables, such as work, heat and entropy \cite{Seifert2005,Jarzynski2011,Sevick2008}. For instance, in the presence of fluctuations, the second law of thermodynamics is only satisfied upon taking suitable averages over an ensemble of stochastic trajectories or over long observation times. 

Over the last decades, the average rate of entropy production, denoted $\dot{S}_i$, has attracted considerable attention as a way of quantifying the degree of departure from equilibrium. For instance, genuinely non-equilibrium processes (as opposed to those relaxing to equilibrium), such as overdamped active particles driven by injection and dissipation of energy at the single-agent level \cite{Bechinger2016,Marchetti2013}, are characterized by a positive average entropy production at steady-state which equals the rate at which heat is dissipated into the environment. 

Interestingly, entropy production has also been formalized as a measure of the breaking of the global detailed balance condition \cite{Cocconi2020, Schnakenberg1976, Lebowitz1999}. In particular, it has long been established for Markovian processes \cite{Gaspard2004} that the thermodynamic entropy production has an equivalent information-theoretic interpretation as the relative dynamical entropy (i.e., the Kullback–Leibler divergence \cite{Kullback1951}) per unit time of the ensemble of forward paths and their time-reversed counterparts, thus signalling the breaking of time-reversal symmetry whenever $\dot{S}_i > 0$. Based on this perspective, it was further shown that the rate of entropy production is inversely proportional to the minimal time needed to decide on the direction of the arrow of time \cite{Roldan2015,Seif2021}. Entropy production has additionally been found to relate non-trivially to the precision and efficiency of the underlying stochastic process via uncertainty relations \cite{Seifert2018,Horowitz2020}. 

In this work, we consider the average entropy production associated with a Brownian particle subject to diffusion in a fluctuating trapping potential $V(x;\alpha(t))$ whose shape is governed by a parameter $\alpha(t)$. In most of what follows, we will assume the potential to be harmonic and centered at the origin, with fluctuations acting solely on the potential stiffness. In the absence of fluctuations, this model reduces to the well-known Ornstein-Uhlenbeck (OU) process \cite{Gardiner1985}, a prototypical equilibrium stochastic process characterized by a Gaussian steady-state probability density function for the particle position $x$, and zero entropy production. As we will demonstrate, letting $\alpha(t)$ evolve stochastically results generically in a departure from thermodynamic equilibrium, signalled by non-vanishing probability currents at steady-state and thus a positive rate of entropy production. 

Introducing fluctuations into what would otherwise be time-independent model parameters is a recurrent theme in non-equilibrium physics. Indeed, think for example of Run-and-Tumble (RnT) and Active Ornstein-Uhlenbeck (AOUPs) particles, whose self-propulsion velocity is described by a telegraph process and an OU process, respectively \cite{Bothe2021,GarciaMillan2021}. Fluctuating interactions are a generic feature of living systems and can have striking consequences including clustering in populations of bacteria interacting via type IV pili \cite{alston2022,Bonazzi2018,Zhou2021}, arrested coalescence in cellular aggregates \cite{Oriola2021} and fluidization of embryonic tissues \cite{Kim2021}. Moreover, a clear thermodynamic understanding of trapping by fluctuating harmonic potentials could have important implications in a number of mesoscopic systems. For instance, experimental manipulation of colloidal beads \cite{Ariga2021} and molecular motor cargoes \cite{Ariga2018,HashemiShabestari2017,Neuman2008,Bustamante2021} by optical tweezers are likely to be subject to non-negligible fluctuations (e.g. from the laser intensity). 

Furthermore, Brownian motion in an intermittent harmonic confining potential represents a realistic implementation of stochastic resetting \cite{Santra2021,Jerez2021,Gupta2020a,Gupta2020b,Evans2020}. Originally introduced to allow Brownian dynamics to reach a nonequilibrium stationary state (NESS) at long times \cite{Evans2011a,Evans2020}, stochastic resetting has been under intense scrutiny over the last decade partly due to its non-trivial impact on first-passage statistics \cite{Evans2011a,Evans2011b} and has imposed itself as a pillar of nonequilibrium statistical mechanics. As a consequence, the effects of resetting have been studied in a swath of physical systems: from classical diffusive processes such as Brownian motion, random walks, L\'{e}vy walks and L\'{e}vy flights \cite{Majumdar2021,Gupta2019,Shkilev2017,Montero2013,Zhou2020,Kusmierz2014}, to the random acceleration process \cite{Singh2020} and the asymmetric exclusion processes \cite{Basu2019,Karthika2020}. More recently, resetting has also found applications in stochastic living systems including in models of active particles \cite{Evans2018a,Santra2020,Kumar2020}, active transport in living cells \cite{Bressloff2020}, enzymatic reactions \cite{Reuveni2014,Reuveni2016}, population genetics \cite{DaSilva2021} and in models of cell division \cite{Genthon2022}.

Of interest here is the fact that the vast majority of these studies generically consider fully irreversible and instantaneous resetting. While various works have addressed the non-equilibrium thermodynamics of resetting, the typically assumed irreversibility of resetting events requires a special treatment \cite{Fuchs2016,Busiello2020,Pal2017}. In particular, these studies made use of alternative definitions for the entropy production whose connection with time-reversal symmetry breaking remains unclear. Here, we argue that a realistic implementation of an effective resetting protocol can offer a relevant perspective on these controversies.

The paper is structured as follows: in Section \ref{sec:derivation}, we derive equations for the steady-state entropy production for a general single-particle drift-diffusion system with fluctuating potentials, considering both discrete and continuous state spaces for the potential states. The rest of the paper is dedicated to specific examples of these single-particle systems. In Section \ref{sec:intermittent}, we consider the simple example of an intermittent harmonic potential to illustrate a practical application of the theory, calculating the steady-state entropy production exactly in Eq. (\ref{eq:SI}). We then consider a generalized two-state OU model in Section \ref{sec:2states} and derive its entropy production in Eq. (\ref{eq:general_si_2state}), before extending this result to an arbitrary number of states in Section \ref{sec:Nstates}, deriving Eq. (\ref{eq:se_gen_N}). In Section \ref{sec:constates}, we study an OU process with a stiffness that varies continuously in time, writing the entropy production in terms of the variance of the particle position in Eq. (\ref{eq:SSEPOUOU}). Finally, our results are summarized in Section \ref{sec:disc}.

\begin{figure}[t]
\centering
\includegraphics[width=0.7\textwidth]{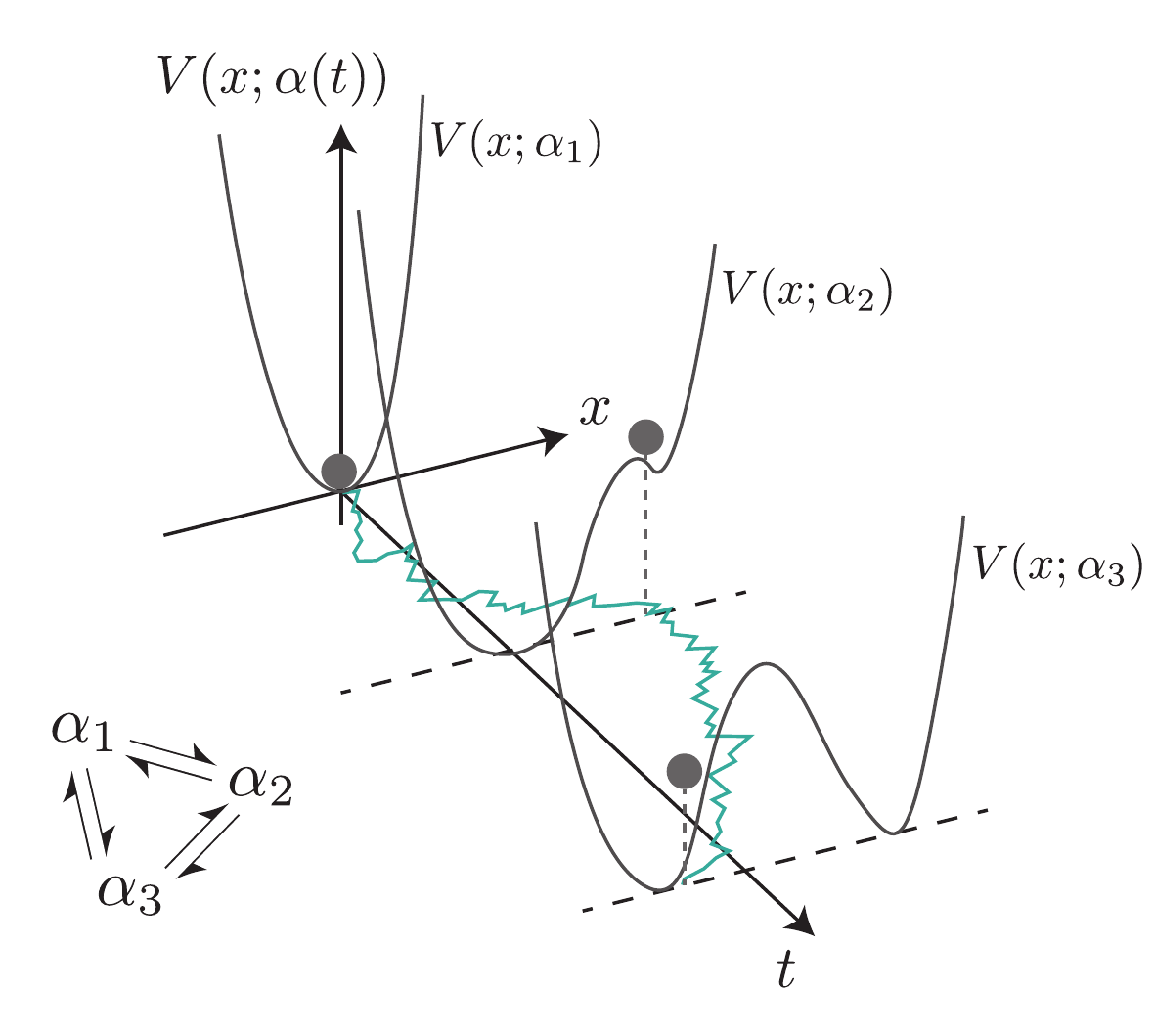}
\caption{{\it Diffusion in fluctuating potentials ---} In many realistic settings, trapping potentials can be subject to stochastic fluctuations. This phenomenon generically breaks global detailed balance and can thus drive a passive Brownian particle trapped in the potential away from thermodynamic equilibrium, such that the corresponding entropy production is non-zero even at steady-state. In the simplest case, a stochastic potential switches between a pre-defined set of functional forms $V_i(x) = V(x;\alpha_i)$, with $i\in \{1,2,3\}$, according to a Markov jump process with given, time-independent transition rates.}
\label{fig:schematic_generic_switch}
\end{figure}

\section{Steady-state entropy production in drift-diffusion processes with fluctuating potentials}
\label{sec:derivation}

In this first section, we derive the general expression for the steady-state entropy production of a Brownian particle diffusing on the real line, $x \in \mathbb{R}$, in a confining potential $V(x;\alpha(t))$, whose shape is set by $\alpha(t)$, a random variable that evolves in continuous time according to Markovian dynamics (see Fig.\,\ref{fig:schematic_generic_switch}). While in the rest of this study we focus on the case of a harmonic confining potential $V(x;\alpha(t)) = \alpha(t) x^2/2$, the functional form of the potential will remain generic in this section. First, we derive the steady-state entropy production in the case where the potential follows a discrete Markov process; in this case, we assume that the potential jumps in between different `states' corresponding to particular values of $\alpha(t)$. We then derive the corresponding results in the case where $\alpha(t)$ follows a generic continuous Markov process.

\subsection{Fluctuating potentials as a discrete Markov process}
\label{sec:derivdiscrete}

First, we model the fluctuations in the confining potential as arising from jumps amongst a finite set of $N$ different states. Namely, we let $\alpha(t) \in \{\alpha_1,\alpha_2,...,\alpha_N\}$ evolve according to a continuous-time, $N$-state Markov jump process with transition rate matrix $K$, where the matrix element $K_{ij}$ with $i \neq j$ denotes the rate at which the parameter switches from value $\alpha_j$ to $\alpha_i$. The diagonal elements are generically fixed by enforcing conservation of total probability, $\sum_{i} K_{ij} = 0$, so that $K_{jj} = - \sum_{i \neq j} K_{ij}$. 

The resulting stochastic dynamics for the parameter $\alpha(t)$ is thus given by
\begin{equation}
	\mathbb{P}\big(\alpha(t+\Delta t)=\alpha_i | \alpha(t)=\alpha_j\big) = \delta_{ij} + \Delta t K_{ij} + \mathcal{O}\big(\Delta t^2\big) 
	\label{eq:alpha_discrete_dyn}
\end{equation}
while the particle position is governed by the overdamped Langevin equation
\begin{equation}
	\dot{x}(t) = -\frac{1}{\gamma}\, \partial_x V(x;\alpha(t)) + \sqrt{2D}\eta(t)~, 
	\label{eq:langevin_general}
\end{equation}
where $\gamma$ is a friction coefficient and $\eta(t)$ denotes a Gaussian white noise with zero mean, $\langle \eta(t)\rangle = 0$, and unit variance, $\langle \eta(t)\eta(t') \rangle=\delta(t-t')$. We set $\gamma=1$ without loss of generality. The corresponding Fokker-Planck equation takes the form \cite{Gardiner1985,Risken1996}
\begin{equation}
	\partial_t P_i(x,t) = -\partial_x J_i(x,t) + \sum_j K_{ij} P_j(x,t) 
	\label{eq:fok_plank_N}
\end{equation}
for $i=1,2,...,N$, with $P_i(x,t)$ the joint probability density that the particle is found at position $x$ with the potential in state $i$ and $J_i(x,t)$ the state-dependent probability current density, given by
\begin{equation}
	J_i(x,t) = -\big(\partial_x V(x;\alpha_i)\big)P_i(x,t) - D \partial_x P_i(x,t)~. 
	\label{eq:curr_def_fp}
\end{equation}
By definition, the total probability is defined as $P(x,t) = \sum_{i=1}^N P_i(x,t)$.

The Gibbs-Shannon entropy \cite{Shannon1948} of the probability density $P(x,t)$ is defined as
\begin{equation}
	S(t) = - \sum_i \int dx \ P_i(x,t) \log \left(\frac{P_i(x,t)}{\bar{P}}\right)
\end{equation}
where $\bar{P}$ is an arbitrary density introduced for dimensional consistency and we work in units such that $k_B = 1$. Differentiating $S(t)$ with respect to time, we see
\begin{equation}
	\dot{S}(t) = - \sum_i \int dx \ \partial_t P_i(x,t) \log \left( \frac{P_i(x,t)}{\bar{P}} \right)
\end{equation}
and using Eq.\,\eqref{eq:fok_plank_N}, we obtain after integration by parts
\begin{equation}
\dot{S}(t) = - \sum_i \int dx \left[ \frac{J_i(x,t) \partial_xP_i(x,t)}{P_i(x,t)} + \sum_j K_{ij} P_j(x,t) \log\left( \frac{P_i(x,t)}{\bar{P}}\right) \right]
\end{equation}
which using Eq.\,\eqref{eq:curr_def_fp}, we rewrite as
\begin{equation}
\dot{S}(t) = \sum_i \int dx \left[ \frac{J_i^2(x,t)}{D P_i(x,t)} + \frac{J_i(x,t) \partial_x V(x;\alpha_i)}{D} - \sum_j K_{ij} P_j(x,t) \log \left(\frac{P_i(x,t)}{\bar{P}} \right)\right].
\label{eq:diff_shann}
\end{equation}

By conservation of probability, we have  
\begin{align}
	K_{ii} P_i(x,t) = -\sum_{j \neq i} K_{ji} P_i(x,t) ~,
\end{align}
which allows us to rewrite the third term on the right-hand side of Eq.~\eqref{eq:diff_shann} as
\begin{align}
	\int dx \sum_{i,j} K_{ij} P_j(x,t)& \log \left(\frac{P_i(x,t)}{\bar{P}}\right) =  \nonumber \\
												 &-\frac{1}{2} \int dx \sum_{i \neq j} (K_{ij} P_j(x,t) - K_{ji}P_i(x,t)) \log \left(\frac{K_{ij} P_j(x,t)}{K_{ji} P_i(x,t)}\right) \nonumber \\
												 &+\frac{1}{2} \int dx \sum_{i \neq j} (K_{ij} P_j(x,t) - K_{ji}P_i(x,t)) \log \left(\frac{K_{ij} }{K_{ji}}\right) ~.
\end{align}

Finally, following the standard procedure \cite{Seifert2012,Cocconi2020}, the contributions to the rate of change of the Gibbs-Shannon entropy are split into two terms 
\begin{equation}
	\dot{S}(t) = \dot{S}_i(t) + \dot{S}_e(t)~,
\end{equation}
with the internal (or total) entropy production defined as
\begin{equation}
	\dot{S}_i(t) = \sum_i \left[ \int dx \ \frac{J_i^2(x,t)}{D P_i(x,t)} \right] + \frac{1}{2} \int dx \sum_{i \neq j} \big(K_{ij} P_j(x,t) - K_{ji}P_i(x,t)\big) \log \left(\frac{K_{ij} P_j(x,t)}{K_{ji} P_i(x,t)}\right)
	\label{eq:en_prod}
\end{equation}
and the external entropy production (or entropy flow) as
\begin{equation}
	\dot{S}_e(t) = \sum_i \left[ \int dx \ \frac{J_i(x,t) \partial_x V(x;\alpha_i)}{D} \right] - \frac{1}{2} \sum_{i \neq j} \big(K_{ij} P_j^{\rm tot}(t) - K_{ji}P_i^{\rm tot}(t)\big) \log \left(\frac{K_{ij} }{K_{ji}}\right),
	\label{eq:en_flow}
\end{equation}
where $P_i^{\rm tot}(t) = \int dx \ P_i(x,t)$ denotes the marginal probability for the potential to be in state $\alpha_i$ at time $t$, irrespective of the particle position. This marginal probability satisfies the master equation
\begin{equation}
	\partial_t P_i^{\rm tot}(t) = \sum_j K_{ij} P_j^{\rm tot}(t) 
	\label{eq:master_eq_alpha}
\end{equation}
and its steady-state value $\lim_{t\to\infty} P_i^{\rm tot}(t)$ can thus be obtained straightforwardly by identifying the unique eigenvector with eigenvalue zero of the matrix $K$. Note that while the entropy flow is commonly associated with the rate of entropy production in the environment \cite{Schnakenberg1976,Lebowitz1999}, the internal entropy production is usually the quantity of interest in the thermodynamic characterization of non-equilibrium stochastic processes due to its connection with time-reversal symmetry breaking \cite{Gaspard2004}, its link to the Kullback-Leibler divergence \cite{Kullback1951} and its role in fluctuation theorems \cite{Seifert2005,Sevick2008}. For the sake of brevity, the denomination of {\it entropy production} will henceforth be reserved for the internal contribution, $\dot{S}_i(t)$, only.
 
Assuming that the joint probability density $P_i(x,t)$ relaxes to a steady-state at long times, we have the equality
\begin{equation}
	\lim_{t \to \infty} \dot{S}(t) =  \lim_{t \to \infty} [ \dot{S}_i(t) + \dot{S}_e(t) ] = 0~.
\end{equation}

While both $\dot{S}_i$ and $\dot{S}_e$ vanish individually only for systems at equilibrium, the internal and external contributions to the entropy production cancel each other exactly even in systems out of thermal equilibrium. As a consequence, the steady-state internal entropy production can equivalently be computed via the entropy flow. This is often a convenient route, since the logarithmic term in Eq.\,\eqref{eq:en_flow} does not contain information about the steady-state distribution itself. 

Note that for Eqs.\,\eqref{eq:en_prod} and \eqref{eq:en_flow} to be well-defined, transitions between potential states $\alpha_i$ must be individually reversible, i.e. $K_{ij} > 0$ if $K_{ji} >0$, while in general $K_{ij} \neq K_{ji}$. If the marginal dynamics for the potential state $\alpha$ satisfy the detailed balance condition \cite{Schnakenberg1976,Lebowitz1999}, i.e. if a global potential function $F_i = F(\alpha_i)$ can be defined such that $K_{ij}/K_{ji} \propto {\rm \exp}(-(F_i - F_j))$ for all pairs $\{i,j\}$, the second term in Eq.\,\eqref{eq:en_flow} vanishes at steady-state, although the first term remains generally positive. This construction is always possible for $N=2$ and, more generally, when the state-space is tree-like, i.e. when it features no closed circuits \cite{Schnakenberg1976}.

\subsection{Fluctuating potentials as a continuous Markov process}
\label{sec:derivcont}

The formulation above can be straightforwardly extended to continuous $\alpha$ dynamics by taking $N \to \infty$ together with a suitable continuum limit in $\alpha$-space, whereby $P_i(x,t) \to P(x,\alpha,t) d\alpha$ and $P_i^{\rm tot}(t) \to P^{\rm tot}(\alpha,t) d\alpha$. In this case, Eq.\,\eqref{eq:alpha_discrete_dyn} thus generalizes to 
\begin{equation}
	\mathbb{P}(\alpha(t+\Delta t)=\alpha' | \alpha(t)=\alpha) = G(\alpha \to \alpha';\Delta t)
\end{equation}
where $G$ denotes the propagator (Green's function) for the chosen dynamics. The associated Fokker-Planck equation, which corresponds to the continuum limit of Eq.~\eqref{eq:master_eq_alpha}, reads
\begin{equation}
	\partial_t P^{\rm tot}(\alpha,t) = \mathcal{L} P^{\rm tot}(\alpha,t)
\end{equation}
with $\mathcal{L}$ the linear Fokker-Planck operator \cite{Risken1996}. For the case of a fluctuating potential with control parameter $\alpha(t)$ described by Brownian motion with diffusion coefficient $D_\alpha$ in a potential $\mathcal{V}(\alpha)$, we have, for instance
\begin{equation}
	\mathcal{L} P^{\rm tot}(\alpha,t)= D_\alpha \partial_\alpha^2 P^{\rm tot}(\alpha,t) + \partial_\alpha( P^{\rm tot}(\alpha,t) \partial_\alpha \mathcal{V}(\alpha) )~. 
	\label{eq:fp_alpha}
\end{equation}
The calculation of the entropy flow starts once again from the expression for the Gibbs-Shannon entropy,
\begin{equation}
	\dot{S}(t) = - \iint dx \ d\alpha \ P(x,\alpha,t) \log \left( \frac{P(x,\alpha,t)}{\bar{P}} \right)~,
\end{equation}
which combines with the now two-dimensional Fokker-Planck equation 
\begin{equation}
	\partial_t P(x,\alpha,t) = - \partial_x J(x,\alpha,t) + \mathcal{L} P(x,\alpha,t)
\end{equation}
to give, for the particular case of Eq.~\eqref{eq:fp_alpha},
\begin{subequations}
\begin{align}
	\dot{S}_i(t) &= \int d\alpha \ dx \ \frac{1}{P(x,\alpha,t)} \left[  \ \frac{J^2(x,\alpha,t)}{D } + \frac{\mathcal{J}^2(x,\alpha,t)}{D_\alpha }  \right]  \label{eq:en_flow_cont_int} \\
	\dot{S}_e(t) &= \int d\alpha \ \left[ \int dx \ \frac{J(x,\alpha,t) \partial_x V(x;\alpha)}{D} \right] 
				+ \frac{1}{D_\alpha} \int d\alpha \ \mathcal{J}^{\rm tot}(\alpha,t) \partial_\alpha  \mathcal{V}(\alpha) \label{eq:en_flow_cont_ext}~.
\end{align}
\label{eq:en_flow_cont}%
\end{subequations}
In line with Eq.~\eqref{eq:curr_def_fp}, the probability current for the particle position satisfies
\begin{equation}
	J(x,\alpha,t) = -(\partial_x V(x;\alpha))P(x,\alpha,t) - D \partial_x P(x,\alpha,t)~,
\end{equation}
while $\mathcal{J}(x,\alpha,t)$ denotes the probability current in $\alpha$-space
\begin{equation}
	\mathcal{J}(x,\alpha,t) = -(\partial_\alpha \mathcal{V}(\alpha))P(x,\alpha,t) - D_\alpha \partial_\alpha P(x,\alpha,t)~,
\end{equation}
with the marginal current $\mathcal{J}^{\rm tot}(\alpha,t) = \int dx \ \mathcal{J}(x,\alpha,t)$.

\section{Brownian motion in an intermittent harmonic potential}
\label{sec:intermittent}

Armed with the general expressions for the entropy production in drift-diffusion processes with fluctuating potentials, we now study a number of specific examples. For the rest of this study, we focus on the case of a harmonic potential $V(x;\alpha(t)) = \alpha(t)x^2/2$, where the fluctuating parameter $\alpha(t)$ controls the potential stiffness. The simplest discrete process that the stiffness of the harmonic potential can follow is a two-state Markov process, also known as dichotomous noise or telegraph process \cite{VanKampen2007}. 

As a preliminary example, we study the case of a fully intermittent harmonic potential \cite{Santra2021}. We suppose that $\alpha(t) \in \{0,\alpha_0\}$ switches between its two states with symmetric rate $k$. The two states are characterized as follows: (i) when $\alpha(t)=0$, the particle diffuses on the real line and we say that the system is in an \textit{off} state, (ii) when $\alpha(t) = \alpha_0 > 0$, the harmonic confining potential is present and the system is said to be in its \textit{on} state. Clearly, in its {\it off} state the particle will be freely diffusing, while in the {\it on} state the confining potential leads to a forcing of the motion of the particle towards the center of the potential. Effectively, this system corresponds to the simplest single-particle system with a non-instantaneous resetting mechanism.

We denote by $P_{\rm off}(x, t)$ and $P_{\rm on}(x, t)$ the joint probability density of finding a particle at position $x$ in the \textit{off} and \textit{on} state, respectively, at time $t$. The kinetic equations for this process read
\begin{subequations}
\begin{align}
	\partial_t P_{\rm off}(x,t) &= -\partial_x \big[J_{\rm off}(x,t)\big] + k P_{\rm on}(x,t) - k P_{\rm off}(x,t) \\
	\partial_t P_{\rm on}(x,t) &= -\partial_x \big[J_{\rm on}(x,t)\big] + kP_{\rm off}(x,t) - kP_{\rm on}(x,t)
\end{align}
\label{eq:kinetic}%
\end{subequations}
with 
\begin{subequations}
\begin{align}
	J_{\rm off}(x,t) &= -D\partial_x P_{\rm off}(x,t), \\
	J_{\rm on}(x,t) &= -D\partial_x P_{\rm on}(x, t) - \alpha_0 x P_{\rm on}(x,t)
\end{align}
\label{eq:fluxintermittent}%
\end{subequations}
The stationary probabilities exist provided that $\alpha_0 > 0$ \cite{Yuan2003}. While it is relatively easy to obtain these stationary probabilities in Fourier space, deriving a closed-form analytic expression for the probability distribution in real space is highly non-trivial \cite{Dubkov2003,Zhang2017,Santra2021} (see also \ref{app:intermittent}). In what follows, we interestingly show that such an analytic form is not required for the calculation of the steady-state entropy production. 

Indeed, to calculate the entropy production for this system, we will evaluate the entropy flow. Starting from Eq.\,\eqref{eq:en_flow}, it is clear that the second term is zero as by construction $K_{{\rm on}, {\rm off}} = K_{{\rm off}, {\rm on}} = k$. For our choice of potentials, the first term reduces to
\begin{equation}
	\dot{S_e}(t) =  \frac{\alpha_0}{D}\int dx  \big[x J_{\rm on}(x,t)\big].
	\label{eq:en_flow_2sI}
\end{equation}
At steady-state, the probability currents satisfy the flux balance equation $\partial_x J_{\rm on}(x,t) = -\partial_x J_{\rm off}(x,t)$. Integrating the right-hand side of Eq.\,\eqref{eq:en_flow_2sI} by parts and substituting one current for the other, we obtain
\begin{equation}
	\lim_{t\rightarrow\infty}\dot{S}_i(t) = \alpha_0\int dx\bigg[\frac{x^2}{2}\partial^2_xP_{\rm off}(x,t)\bigg].
	\label{eq:EP2SOUMS}
\end{equation}
Finally, integrating by parts twice leaves us with the simple expression
\begin{equation}
	\lim_{t\rightarrow\infty}\dot{S}_i(t) =\frac{\alpha_0}{2}~,
	\label{eq:SI}
\end{equation}
indicating that the steady-state entropy production in this setup is independent of both the switching rate $k$ and the diffusion coefficient $D$. Here and in the following, we drop boundary terms whenever integration by parts is performed. This procedure relies on a sufficiently fast decay of the relevant probability densities as $x \to \pm\infty$ and, more precisely, on the finiteness of the second moment of $P_i(x)$, which is a reasonable assumption for all processes considered herein. 

\begin{figure}[t!]
\centering
\includegraphics[width=\textwidth]{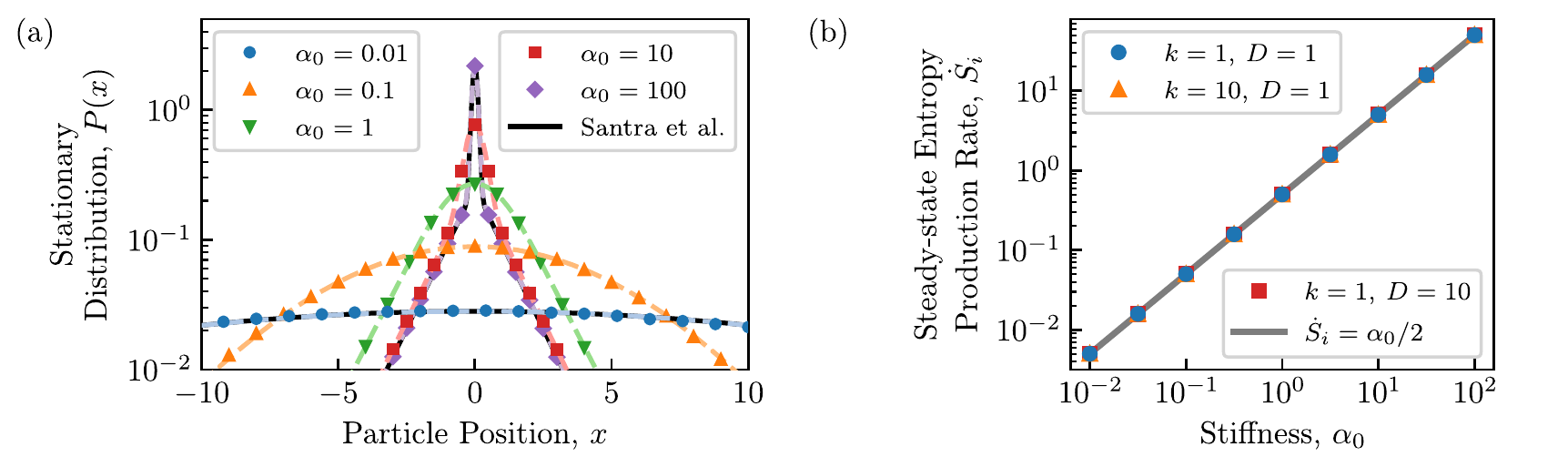}
\caption{{\it Steady-state entropy production of a Brownian particle in an intermittent quadratic potential ---} (a) Stationary distribution $P(x)$ at different values of $\alpha_0\in[10^{-2}, 10^2]$ with $k=D=1$ fixed. We show agreement between the distributions measured numerically from single particle trajectories (marked by symbols) and the result \eqref{eq:real_conv_distribution} which we have integrated numerically (dashed lines). We plot in black the analytic solutions for the limit $\alpha_0\gg k$ as in Eq.\eqref{eq:ss_intermittent_asymptotic_1} and $\alpha_0\ll k$ as in Eq.\eqref{eq:ss_intermittent_asymptotic_2}. (b) We confirm our analytic result \eqref{eq:SI} by evaluating \eqref{eq:en_flow_2sI} numerically from our stationary distributions for three sets of values for $k, D$. 	
}
\label{fig:intermittent}
\end{figure}

As shown in Fig.\,\ref{fig:intermittent}, we confirm numerically this result through: (1) the numerical integration of Eq.\,\eqref{eq:en_flow_2sI} using the stationary current derived from (see \ref{app:intermittent} and \cite{Santra2021,Zhang2017}) and {(2) the analysis of single particle trajectories from the simulated underlying microscopic process governed by Eq.\,\eqref{eq:langevin_general} (see \ref{app:numerical} for further numerical details)}. Strikingly, while the process is \emph{dynamically} equivalent in the limit $k \to \infty$ to an equilibrium OU process with reduced potential stiffness $\alpha_0/2$ \cite{Pavliotis2008}, we observe here a finite and strictly positive steady-state rate of entropy production. Similarly, entropy production remains finite and positive for free run-and-tumble particles, whose motion is effectively diffusive in the limit of infinite tumbling rate or large times \cite{Cocconi2020,GarciaMillan2021}. This is sometimes referred to as an \emph{entropic anomaly} \cite{Celani2012,bo2014entropy}.

Note that the independence of the steady-state entropy production vis-\`{a}-vis the switching rate $k$ and the diffusion coefficient $D$ is specific to our choice of potential and can be derived from physical arguments. Namely, the first law of thermodynamics at steady-state,
\begin{equation}
	0 = \int dx \left[ V(x; \alpha_0)\partial_t P_{\rm on}(x,t) + V(x; 0)\partial_t P_{\rm off}(x,t) \right] = \dot{W} - \dot{Q}~,
\end{equation}
imposes the rate of heat dissipation, $\dot{Q}$, to be equal to the work done per unit time by the potential on the particle, $\dot{W}$. Clearly, work is only being done in the \textit{on} state as the potential disappears in the \textit{off} state. In turn, the average work done equals the change in average potential energy $U = \langle \alpha_0 x^2/2 \rangle$ before the next transition to the \textit{off} state. In the \textit{off} state, the particle motion is purely diffusive and the variance of the position probability density grows linearly, i.e. $\partial_t \langle x^2 \rangle = 2D$. Thus, the average work done by the potential during a \textit{on} phase of typical duration $k^{-1}$ is given by
\begin{equation}
	\langle W \rangle = \frac{D \alpha_0}{k}~.
	\label{eq:WorkDone}
\end{equation}
Given that the average duration of an \textit{on-off} cycle is by construction $2/k$, the average rate of heat dissipation is $\langle \dot{Q} \rangle= k\langle W \rangle/2 = D\alpha_0/2.$ Finally, the particle self-diffusion coefficient being proportional to the temperature by Einstein's relation, we write in our units that $\dot{S}= \langle \dot{Q} \rangle / D$ and finally recover $\dot{S}= \alpha_0 / 2$, which we confirm to be independent of $k$ and $D$.

Importantly, this argument relies on the variance $\langle x^2 \rangle$ growing linearly with time (without bounds) in the \textit{off} state, an assumption that breaks down as soon as the \textit{off} state of the potential has a finite stiffness $\alpha_1$, in which case the variance of the particle position in the \textit{off} state instead satisfies $\partial_t \langle x^2 \rangle = 2D (1-\alpha_1\langle x^2 \rangle/D )$. We show how this leads to an explicit $k$ dependence of the entropy production in Section \ref{sec:non_vanish_alpha}. Finally, to highlight the importance of the functional form of the potential, we repeat this procedure for an intermittent \textit{quartic} potential and argue that the steady-state entropy production can not be independent of $k$ or $D$ in \ref{app:quartic}.

\section{General two-state Ornstein-Uhlenbeck Markov process}
\label{sec:2states}

\begin{figure}[t!]
\centering
\includegraphics[width=\textwidth]{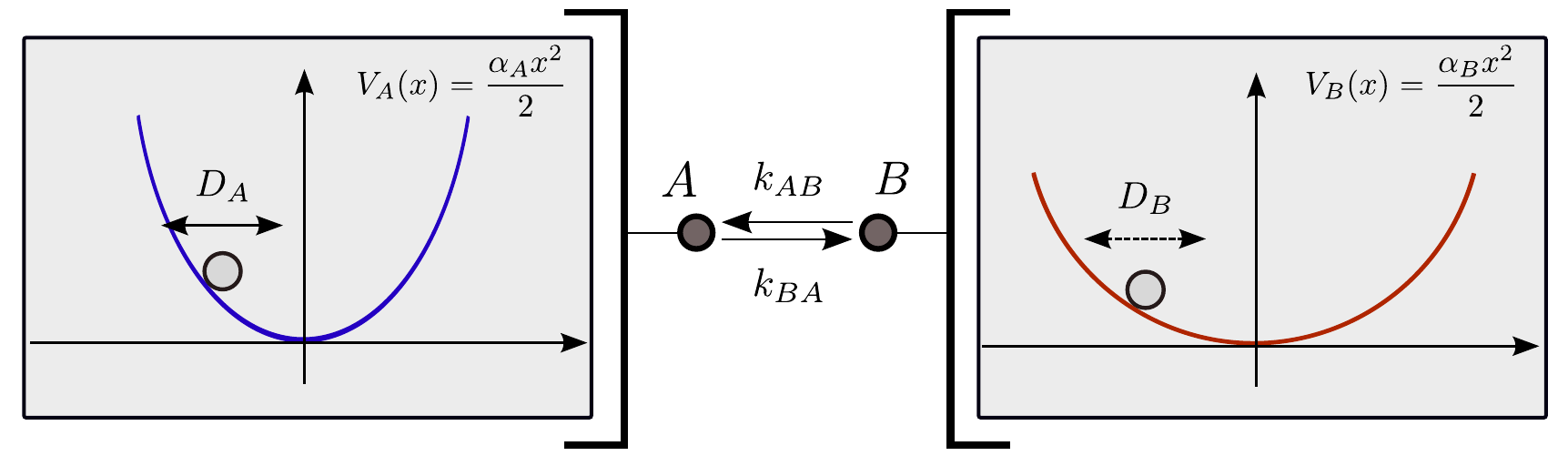}
\caption{{\it Fluctuating potentials as a general two-state Ornstein-Uhlenbeck Markov process ---} Note that one can only obtain the steady-state entropy production when the stationary probabilities exist; for this system, the inequality \eqref{eq:SPexist} must be satisfied.}
\label{fig:schematic}
\end{figure}

We now broaden our focus and study the case of a generalized two-state Ornstein-Uhlenbeck Markov process, of which the preliminary model introduced in the previous section is a limiting case. Here, we consider a system with two states denoted $A$ and $B$. In state $A$, the particle diffuses in a harmonic potential, $V_A(x) = \alpha_A x^2 / 2$, with diffusion coefficient $D_A$. In state $B$, the restoring force comes from a second potential, $V_B(x) = \alpha_B x^2 / 2$, and the particle self-diffusion is set by $D_B$. As shown in Fig.~\ref{fig:schematic}, the particle switches from state $A$ to $B$ with rate $k_{BA}$ and returns with rate $k_{AB}$. 

\subsection{Analytic expression for the entropy production}
\label{sec:g2statesEP}

In this general case, one needs to carefully chose the potential strengths and switching rates. Indeed, the steady-state probabilities only exist for this system when the following inequality is satisfied \cite{Yuan2003}:
\begin{equation}
	\frac{\alpha_A k_{AB} + \alpha_B k_{BA}}{k_{AB}+k_{BA}} = \langle \alpha \rangle > 0.
	\label{eq:SPexist}
\end{equation}
Namely, while the independent confining potential strengths do not need to be strictly positive, we require the effective potential strength (as time-averaged over a full $A \to B \to A$ cycle) to be positive.

Granted that condition \eqref{eq:SPexist} is met, we start from Eq.\,\eqref{eq:en_flow} and follow the same procedure as above. We thus argue that the steady-state entropy flow reads
\begin{equation}
	\lim_{t\rightarrow\infty}\dot{S}_e(t) = \frac{\alpha_A}{D_A} \int dx \big[x J_A(x)\big]+\frac{\alpha_B}{D_B} \int dx \big[x J_B(x)\big].
\end{equation}
 We then substitute in the form of the currents from the Fokker-Planck equations for the process, given in Eq.~\eqref{eq:curr_def_fp}, and write the internal entropy production as 
\begin{equation}
	\lim_{t\rightarrow\infty}\dot{S}_i(t) = -\langle \alpha \rangle +\frac{\alpha_A^2}{D_A}\int dx \big[x^2 P_A(x)\big] + \frac{\alpha_B^2}{D_B}\int dx \big[x^2 P_B(x)\big],
	\label{eq:EPG2OUMS}
\end{equation}
where we recognize that the two integrals are proportional to the variances of the steady-state probability distributions conditioned on the potential being in either of the two states $A$ and $B$. 

We introduce the conditional variance
\begin{equation}
\sigma_i^2(t) = \frac{\int dx \ x^2 P_i(x,t)}{\int dx \ P_i(x,t)}
\end{equation}
and define 
\begin{equation}
	\Xi_i(t) = \int dx \ x^2 P_i(x,t) = \sigma_i^2(t) P_i^{\rm tot} (t)
	\label{eq:Xi_def}
\end{equation}
for $i \in \{A,B\}$, where $P_i^{\rm tot}(t) = \int dx \ P_i(x,t)$ is the marginal probability of the potential having stiffness $\alpha_i$ independent of the position $x$ of the trapped particle. 

First, note that 
\begin{equation}
	\partial_t\Xi_i(t)  =\int dx \big[x^2 \partial_t P_i(x,t)\big]
\end{equation}
and so after taking the second moment of the Fokker-Planck equation \eqref{eq:fok_plank_N}, we obtain
\begin{subequations}
\begin{align}
	\partial_t\Xi_A(t)  & = \frac{2 D_A k_{AB}}{k_{AB} + k_{BA}} -(2\alpha_A+k_{BA})\Xi_A(t)  + k_{AB}\Xi_B(t)  \label{eq:V0G2OUMS} \\
	\partial_t\Xi_B(t)  & = \frac{2 D_B k_{BA}}{k_{AB} + k_{BA}} -(2\alpha_B+k_{AB})\Xi_B(t)  + k_{BA}\Xi_A(t) . \label{eq:V1G2OUMS}
\end{align}
\label{eq:G2OUMS}
\end{subequations}

We can now solve Eqs.\,\eqref{eq:V0G2OUMS} and \eqref{eq:V1G2OUMS} at steady-state to derive explicit expressions for $\Xi_A(t)$ and $\Xi_B(t)$ as $t\rightarrow\infty$:
\begin{subequations}
\begin{align}
\lim_{t\rightarrow\infty} \Xi_A(t) &= \frac{k_{AB}}{k_{AB}+k_{BA}}\bigg[ \frac{(2\alpha_B + k_{AB})D_A + k_{BA} D_B}{2\alpha_A\alpha_B + \alpha_Ak_{AB} + \alpha_Bk_{BA}}\bigg], \label{eq:VG2OUMSa} \\
\lim_{t\rightarrow\infty} \Xi_B(t) &= \frac{k_{BA}}{k_{AB}+k_{BA}} \bigg[\frac{k_{AB} D_A + (2\alpha_A + k_{BA})D_B}{2\alpha_A\alpha_B + \alpha_Ak_{AB} + \alpha_Bk_{BA}}\bigg]. \label{eq:VG2OUMSb} 
\end{align}
\end{subequations}
Substituting \eqref{eq:VG2OUMSa} and \eqref{eq:VG2OUMSb} into \eqref{eq:EPG2OUMS}, we obtain a closed-form exact expression for the entropy production in a general two-state Ornstein-Uhlenbeck Markov process, which reads
\begin{align}
\lim_{t\rightarrow\infty}\dot{S}_i(t) = &-\frac{\alpha_A k_{AB} + \alpha_B k_{BA}}{k_{AB}+k_{BA}} \nonumber \\
						      &+\frac{\alpha_A^2}{D_A}\left[\frac{k_{AB}}{k_{AB} + k_{BA}}\right]\left[\frac{(2\alpha_B + k_{AB})D_A +  k_{BA} D_B }{2\alpha_A\alpha_B + \alpha_Ak_{AB} + \alpha_Bk_{BA}}\right]\nonumber\\
						      &+ \frac{\alpha_B^2}{D_B}\left[\frac{k_{BA}}{k_{AB} + k_{BA}}\right]\left[\frac{k_{AB}D_A + (2\alpha_A + k_{BA})D_B}{2\alpha_A\alpha_B + \alpha_Ak_{AB} + \alpha_Bk_{BA}}\right]~.\label{eq:general_si_2state}
\end{align}

\subsection{Some models of interest}
\label{sec:modelsinterest}

We now apply this result to a number of important limiting cases of the generalized two-state Ornstein-Uhlenbeck model. 

\subsubsection{Intermittent Harmonic Potential with Asymmetric Switching Rates ---}
\label{sec:asymmintermittent}
First we return to the preliminary example, in which we stipulated that the diffusion was independent of the state, $D_A = D_B = D$, and we let $\alpha_A=0$ and $\alpha_B=\alpha_0$. Here, we consider more generally the case of distinct switching rates: $k_{\rm{on}}$ to switch from state $A$ to state $B$ and $k_{\rm{off}}$ from state $B$ to $A$. For these parameters, Eq.~\eqref{eq:general_si_2state} reduces to
\begin{equation}
	\lim_{t\rightarrow\infty}\dot{S}_i(t) = \alpha_0 \frac{k_{\rm off}}{k_{\rm on} + k_{\rm off}} =\alpha_0 - \langle \alpha\rangle.
	\label{eq:assym}
\end{equation}

We conclude that in this case, the entropy production explicitly depends on the switching rates $k_{\rm{on}}$ and $k_{\rm{off}}$. We interpret the RHS of Eq.~\eqref{eq:assym} as the effective confinement strength as weighted by the fraction of the time the confining potential is {\it off}. Note that we naturally recover the result from Eq.\,\eqref{eq:SI} when symmetrizing the switching rates and setting $k_{\rm{on}} = k_{\rm{off}}$.

\begin{figure}[t!]
	\centering
	\includegraphics[width=\textwidth]{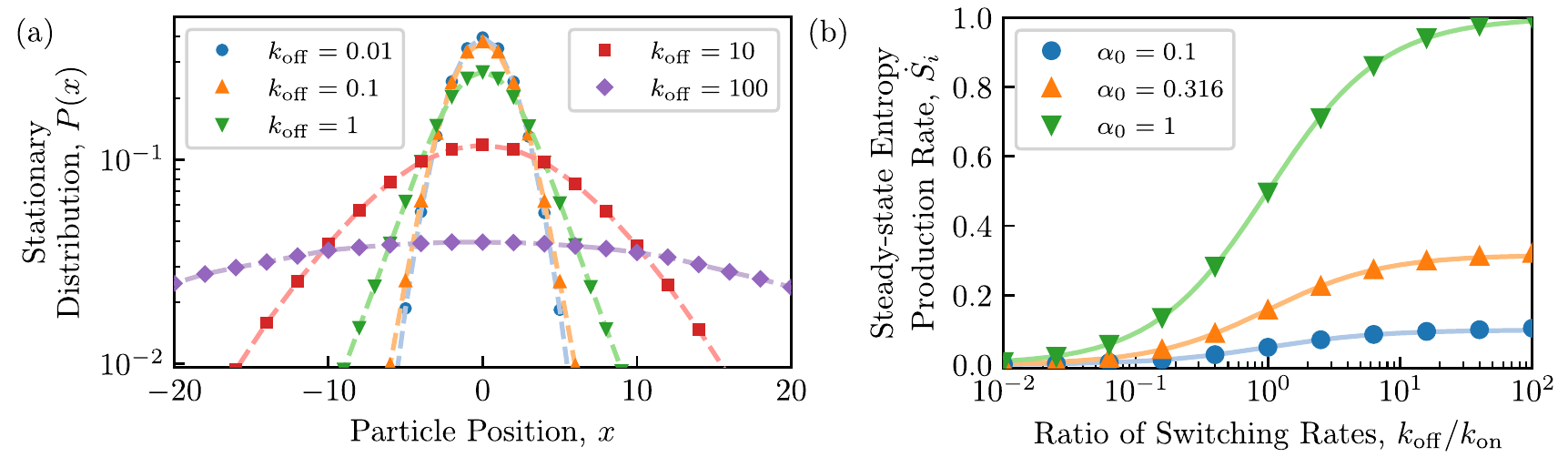}
	\caption{{\it Steady-state entropy production rate for a Brownian particle in an intermittent quadratic potential with asymmetric switching rates ---} (a) Stationary distribution $P(x)$ for the particle position measured numerically from single particle trajectories for varying switching rates $k_{\rm off}\in\{10^{-2}, 10^2\}$ with $k_{\rm on}=D=\alpha_0 = 1$ fixed. (b) Entropy production rate measured by integrating \eqref{eq:EPG2OUMS} numerically (symbols) showing good agreement with our analytic result, \eqref{eq:assym}, for fixed $k_{\rm on} = D = 1$.}
	\label{fig:intermittentasym}
\end{figure}

\subsubsection{Non-disappearing harmonic potential ---}
\label{sec:non_vanish_alpha}

\begin{figure}[b]
\centering
\includegraphics[width=\textwidth]{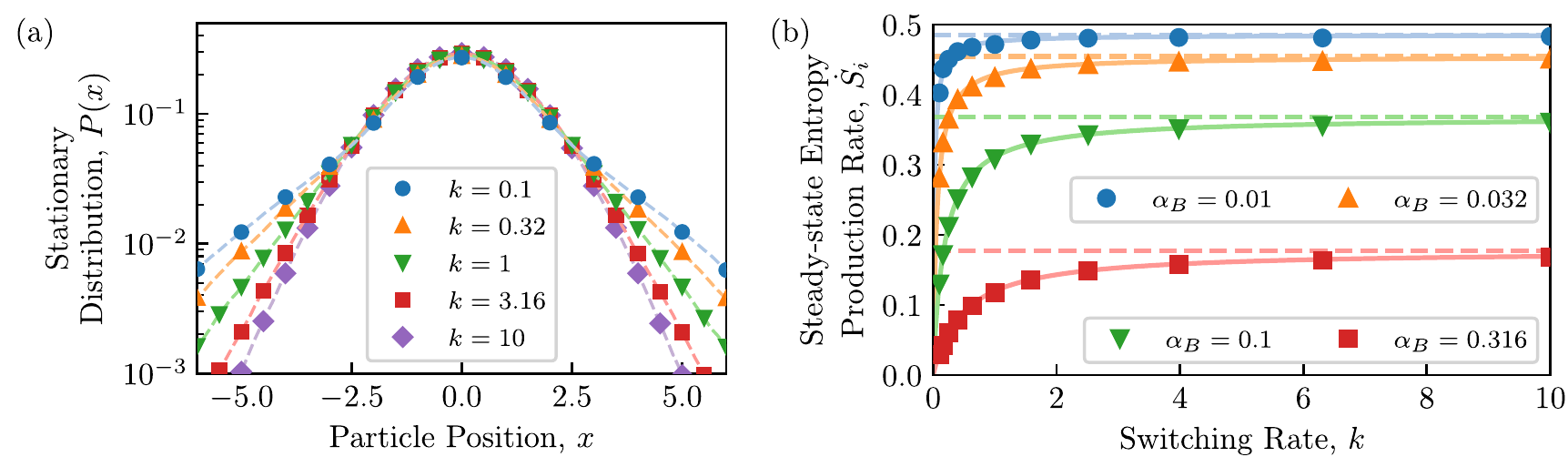}
\caption{{\it Steady-state entropy production rate for a Brownian particle in a harmonic potential switching between two non-zero stiffnesses with rate $k$ ---} (a) Stationary distributions for the process with $\alpha_B = 0.1$ and $\alpha_A=D=1$ for switching rates $k\in\{0.1, 10\}$. (b) Entropy production rate evaluated from the numerical integration of \eqref{eq:EPG2OUMS} using the stationary distributions obtained from single particle trajectories (symbols). We show a perfect quantitative agreement with our exact analytical result \eqref{eq:nonvanEP} (solid line) for a wide range of switching rates. We also show the entropy production rate in the limit $k\rightarrow \infty$ in each case from \eqref{eq:kinf} (dashed line). }
\label{fig:k_dependence_nonvanishing}
\end{figure}

Next, we consider the case of a non-disappearing harmonic potential. Namely, we consider that $k_{AB}=k_{BA}=k$ and $D_A = D_B = D$, while letting $\alpha_A > \alpha_B > 0$. Here, we obtain
\begin{equation}\label{eq:nonvanEP}
	\lim_{t\rightarrow\infty}\dot{S}_i(t) = -\frac{\alpha_A + \alpha_B}{2} + \frac{(\alpha_B+k)\alpha_A^2 + (\alpha_B+k) \alpha_B^2}{2 \alpha_A \alpha_B + k(\alpha_A + \alpha_B)} = \frac{k(\alpha_A - \alpha_B)^2}{4\alpha_A\alpha_B + 2 k (\alpha_A + \alpha_B)}~,
\end{equation}
which displays an explicit dependence on the switching rate $k$. In particular, we observe a crossover from a small $k$ regime characterized by a linear $k$ dependence 
\begin{equation}
	\lim_{t\rightarrow\infty}\dot{S}_i(t) \simeq \frac{(\alpha_A - \alpha_B)^2}{4\alpha_A \alpha_B}k \quad {\rm for} \quad k \ll \alpha_{A,B}
\end{equation}
that extends from $k=0$ up to a cross-over rate $k^* = 2\alpha_A \alpha_B/(\alpha_A + \alpha_B)$, to a large $k$ regime that is asymptotically independent of $k$,
\begin{equation}\label{eq:kinf}
	\lim_{k\rightarrow\infty}\lim_{t\rightarrow\infty}\dot{S}_i(t) = \frac{(\alpha_A - \alpha_B)^2}{2(\alpha_A + \alpha_B)}~.
\end{equation}
We conclude that the $k$-dependent regime vanishes to a single point in the limit where $\alpha_B \to 0$ for fixed $\alpha_A$ as shown in Fig.\,\ref{fig:k_dependence_nonvanishing}. This limit is consistent with a vanishing intermittent harmonic potential and we confirm here that we recover the result of Eq.\,\eqref{eq:SI}.

\subsubsection{Switching diffusion in harmonic potential ---}
\begin{figure}[b]
\centering
\includegraphics[width=\textwidth]{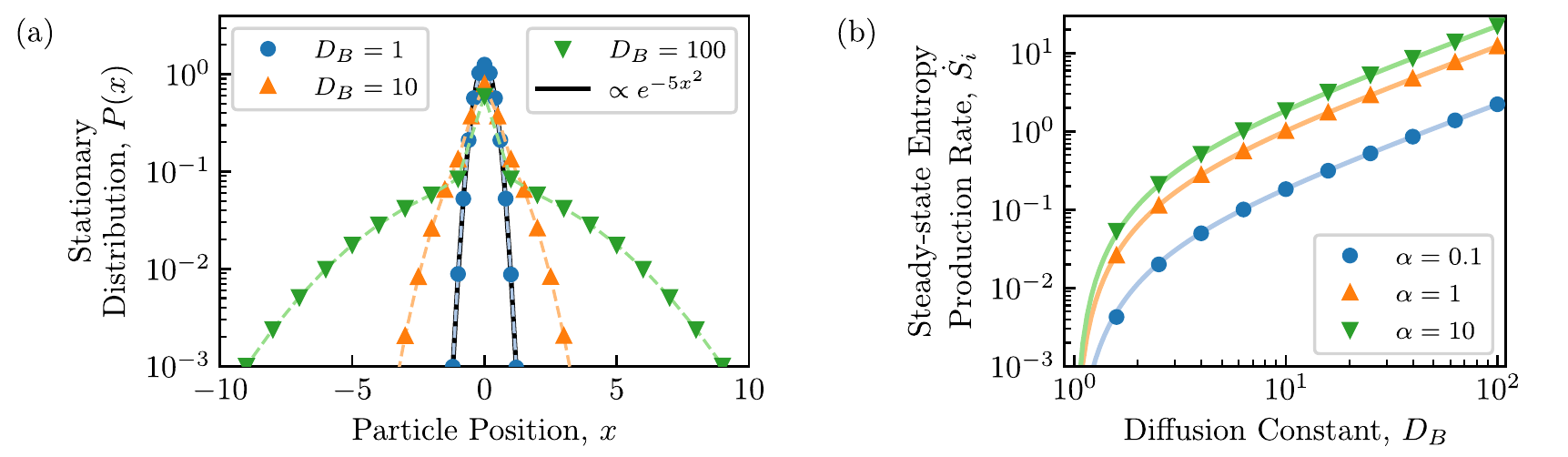}
\caption{{\it Steady-state entropy production for a Brownian particle switching between two diffusion coefficients in a constant harmonic potential ---} (a) Stationary distributions for the process with $D_A = k = 1$, $\alpha=10$ and varying $D_B\in[1, 100]$. We show in black the Gaussian distribution expected in the case where $D_A = D_B $. (b) Entropy production rate as a function of the diffusion coefficient $D_B$ obtained by integrating numerically \eqref{eq:EPG2OUMS} using the single particle trajectories (symbols), in perfect agreement with our analytic result \eqref{eq:2dif} (solid line).  }
\label{fig:SDEP}
\end{figure}

Suppose now that the switching is symmetric with rate $k_{AB}=k_{BA}=k$ and the harmonic potential stiffness $\alpha_A = \alpha_B = \alpha$ is the same in each state, but the diffusion coefficient switches between two values, $D_A$ and $D_B$. We then vary $D_A$ and $D_B$ to see how the entropy production depends on the ratio of the diffusion coefficients. Starting from \eref{eq:general_si_2state}, we eventually obtain
\begin{equation}\label{eq:2dif}
	\lim_{t \to \infty} \dot{S}_i(t) = \frac{\alpha k}{4(\alpha + k)}\left[ \left( \frac{D_B}{D_A} + \frac{D_A}{D_B} \right) -2 \right] = \frac{\alpha k (D_A-D_B)^2}{4 D_AD_B(\alpha + k)}~.
\end{equation}
The entropy production is clearly non-negative and vanishes at $D_A = D_B$, which corresponds to the recovery of a standard (equilibrium) Ornstein-Uhlenbeck process with stiffness $\alpha$ and diffusion coefficient $D$ (see Fig.\,\ref{fig:SDEP} for a comparison with numerical results).

\subsubsection{Effective resetting with harmonic potential ---}
\label{sec:reset}
 
Evaluating the entropy production for systems with (instantaneous) resetting is a problem that has seen much attention \cite{Fuchs2016,Busiello2020}. The irreversible nature of the stochastic resetting process is a good indication that the entropy production is infinite: it completely breaks time-reversal symmetry. For this reason, previous work addressing the thermodynamics of resetting has made use of alternative definitions for the entropy production \cite{Fuchs2016,Busiello2020}, whose connection with time-reversal symmetry breaking in the spirit of \cite{Gaspard2004} is unclear. 

The framework we introduce here allows us to study models of \textit{effective} resetting, where a particle diffuses in a fluctuating harmonic potential. Namely, near-instantaneous resetting with a refractory period \cite{Evans2018b} of typical duration $1/k_{\rm off}$ can be modeled with an intermittent potential of infinite stiffness $\alpha_0 \to \infty$. Note that in the limit where $k_{\rm off} \to \infty$ while keeping $k_{\rm off} \ll \alpha_0$, this refractory period vanishes. From the results of Section\,\ref{sec:asymmintermittent}, it is clear that an infinitely stiff confining potential implies infinite steady-state entropy production, but more generally, we are here able to quantify entropy production in systems approaching instantaneous resetting but with finite confining potentials and show that the entropy production diverges linearly with the potential stiffness.

\section{General $N$-state Ornstein-Uhlenbeck Markov process}
\label{sec:Nstates}

\subsection{General framework}
\label{sec:nstategeneral}

We generalize the results above to the case where the stiffness $\alpha$ of the confining harmonic potential can switch stochastically between $N$ distinct values $\alpha_i$ with $i=1,2,...,N$ following a general Markov jump process with transition rate matrix $K$. As noted earlier, the matrix elements $K_{ij}$ represent the probability per unit time that a harmonic potential with stiffness $\alpha_j$ switches to stiffness $\alpha_i$; note that in general, $K_{ij} \neq K_{ji}$. The diagonal elements of $K$ are fixed by imposing $\sum_i K_{ij} = 0$ for all $j$, corresponding to the requirement that the total probability be conserved. In the following, $P_i(x)$ will denote the joint probability of finding a particle at position $x$ while the potential has stiffness $\alpha_i$. 
Similarly to Eq.\,\eqref{eq:Xi_def}, we define
\begin{equation}
	\Xi_i(t) = \int dx \ x^2 P_i(x,t) = \sigma^2_i(t) P_i^{\rm tot}(t)
\end{equation}
for $i=1,2,...,N$, with $P_i^{\rm tot}(t) = \int dx \ P_i(x,t)$ the marginal probability of the potential having stiffness $\alpha_i$ independently of the particle position. For our choice of potential, the entropy flow is given by Eq.\,(\ref{eq:en_flow}):
\begin{equation}
	\dot{S}_e(t) = \langle \alpha \rangle - \sum_i \frac{\alpha_i^2}{D_i} \Xi_i(t) - \frac{1}{2} \sum_{i\ne j} \big(K_{ij} P_j^{\rm tot}(t) - K_{ji} P_i^{\rm tot}(t)\big) \log\left( \frac{K_{ij}}{K_{ji}}\right). 
	\label{eq:se_gen_N}
\end{equation}
where $\langle \alpha \rangle$ is the mean stiffness
\begin{equation}
	\langle \alpha \rangle = \sum_i \alpha_i P_i^{\rm tot}~.
\end{equation}
Note that in the present case, the contribution from the pure switching component of the process does not generally vanish for $N \geq 3$, since the switching rates $K_{ij}$ do not generically need to satisfy the detailed balance condition. Based on the Fokker-Planck equation \eqref{eq:fok_plank_N}, we obtain the following system of kinetic equations
\begin{equation}
	\partial_t \Xi_i(t) = 2 D_i P_i^{\rm tot} - 2 \alpha_i \Xi_i(t) + \sum_j K_{ij} \Xi_j(t)~.
	\label{eq:kineticeq_Xi_Nstate}
\end{equation}
At steady-state, computing the entropy production $\lim_{t\to \infty} \dot{S}_i(t) = - \lim_{t\to \infty} \dot{S}_e(t)$ for this system only depends on our ability to compute the quantities $\Xi_i$ at steady-state. From Eq.\,\eqref{eq:kineticeq_Xi_Nstate}, these steady-state quantities can be obtained by solving the linear system
\begin{equation}
	2 D_i P_i^{\rm tot} + \sum_j(K_{ij} - 2\alpha_i \delta_{ij}) \Xi_j = 0 
	\label{eq:steady_Xi_N}
\end{equation}
which involves the steady-state marginal probabilities $P_i^{\rm tot}$ of the Markov switch process; these correspond to the unique eigenvector with eigenvalue 0 of the transition rate matrix $K$ (assuming that the corresponding graph has a single connected component), rather than the full space-dependent probabilities $P_i(x,t)$, which are typically hard to compute \cite{Zhang2017,Santra2021}. 

\subsection{Simple example: $N=3$ with homogeneous right- and left-hopping rates}
\label{sec:nstateexample}

\begin{figure}
\centering
\includegraphics[width=\textwidth]{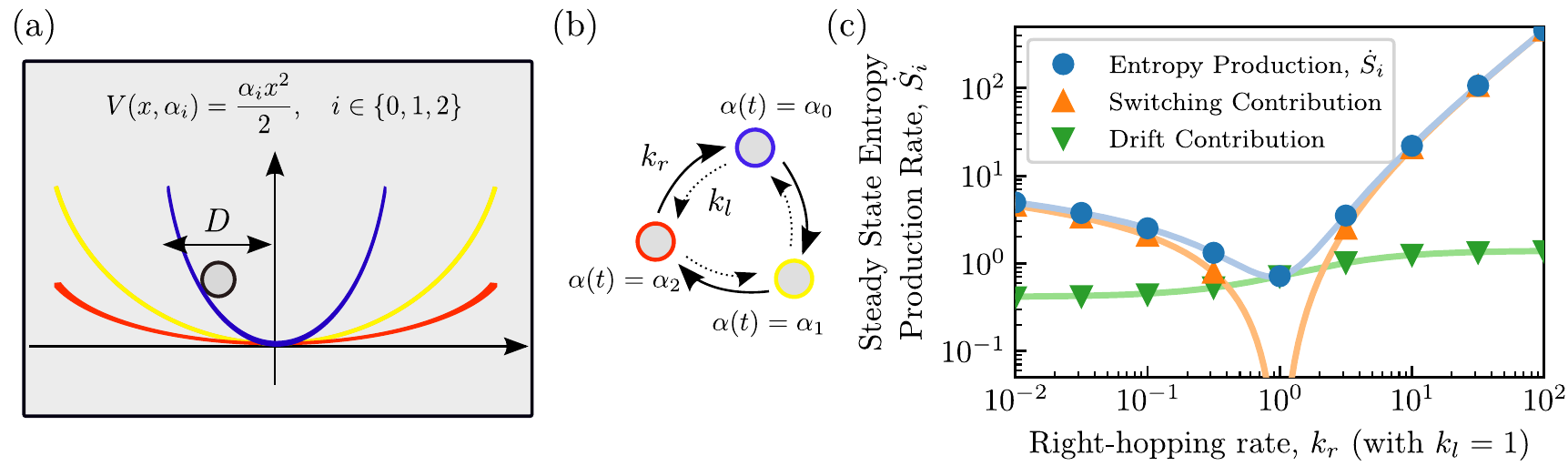}
\caption{{\it Steady-state entropy production for a Brownian particle in $N=3$ harmonic potentials of varying stiffness ---} Schematic for (a) the different potentials in a typical three state system and (b) the discrete Markov jump process that controls the stiffness of the harmonic potential. (c)	Entropy production rates for the process, consisting of two contributions: the {\it switching} contribution, originating purely from the switching dynamics, vanishes when the jump process satisfies detailed balance (here, $k_r = k_l$); the {\it drift} contribution, accounting for steady-state currents in position space, is generically positive in the presence of stiffness fluctuations. For the simulations, we set $\alpha_0 = 0.5,\: \alpha_1 = 2,\:\alpha_2 = 5$ and $D=k_l=1$.
}
\label{fig:N3_switch}
\end{figure}

The case of an $N=3$ ring of states (see Fig.\,\ref{fig:N3_switch}(a)) is the simplest setup for which the switching contribution to the entropy flow is non-trivial. Let $K_{i,i+1} = k_l$ and $K_{i,i-1} = k_r$ with periodic boundary conditions and assume that in all states, particles have the same self-diffusivity, $D_1 = D_2 = D_3 = D$. By rotational symmetry of the ring of states we have $P^{\rm tot}_i = 1/3$ for all $i$. We thus also have that $\langle \alpha \rangle = (\alpha_1 + \alpha_2 + \alpha_3)/3$. The contribution to the entropy flow from the switching part of the process can easily be calculated and reads
\begin{equation}
\frac{1}{2} \sum_{i\ne j} \big(K_{ij} P_j^{\rm tot}(t) - K_{ji} P_i^{\rm tot}(t)\big) \log\left( \frac{K_{ij}}{K_{ji}}\right) = (k_l - k_r) \log \left( \frac{k_l}{k_r}\right)
\end{equation}
which vanishes for $k_l = k_r$, as expected. We can now plug this result into Eq.\,\eqref{eq:se_gen_N} to obtain an expression for the entropy production as a function of the stiffnesses $\alpha_i$ and the switching rates,
\begin{equation}
\lim_{t\to \infty}\dot{S}_i(t) = \lim_{t\to\infty}-\dot{S}_e(t) = (k_r - k_l) \log \frac{k_r}{k_l} + \sum_{i=1}^3 \alpha_i \left( \frac{\alpha_i \Xi_i}{D} - \frac{1}{3}\right) .
\label{eq:s_e_3switch}
\end{equation}
where the steady-state quantities $\Xi_i$ are solutions to the following linear equation 
\begin{equation}
\frac{2D}{3} \mathbf{1} + \bm{\Sigma} \cdot \bm{\Xi} = \mathbf{0}
\end{equation}
with the matrix $\bm{\Sigma}$ defined as 
\begin{equation}
\bm{\Sigma} =  \begin{pmatrix}
				-k_r-k_l-2\alpha_1 & k_l & k_r \\
				k_r  & -k_r-k_l-2\alpha_2 & k_l \\
				k_l & k_r & -k_r-k_l-2\alpha_3\\
		   \end{pmatrix}
\end{equation}
We can thus obtain them by a simple matrix inversion
\begin{equation}
\bm{\Xi} = - \frac{2D }{3}\bm{\Sigma}^{-1} \cdot \mathbf{1}
\end{equation}
From Eq.\,\eqref{eq:s_e_3switch}, we note that the entropy production in this general $N$-state Markov jump process is formed of two contributions: (i) the first term, which we call {\it switching} contribution, originates purely from the switching dynamics and vanishes when the jump process satisfies detailed balance (here, $k_r = k_l$); (ii) the second term, which we call {\it drift} contribution, accounts for steady-state currents in position space and is generically positive in the presence of stiffness fluctuations. 

In the special case of equal stiffnesses $\alpha_1 = \alpha_2 = \alpha_3 = \alpha$, we obtain $\Xi_i = D/(3\alpha)$ and the {\it drift} contribution in Eq.\,\eqref{eq:s_e_3switch} vanishes. In contrast, Figure \ref{fig:N3_switch} shows the evolution of the entropy production as a function of $k_r/k_l$ for a more general case where the stiffnesses $\alpha_i$ are not equal.

\section{Continuous state Markov process for the potential stiffness}
\label{sec:constates}

The final generalization consists in allowing the potential stiffness $\alpha$ to vary continuously according to a continuous stochastic process. While we derived general results about the internal entropy production and the entropy flow for continuous processes in Section\,\ref{sec:derivcont}, here we focus on the particular example of a particle diffusing in a confining harmonic potential whose stiffness obeys an Ornstein-Uhlenbeck process.

\subsection{Ornstein-Uhlenbeck process governing the potential stiffness}

In this model, the position of the particle $x$ obeys the following overdamped Langevin equation 
\begin{equation}
\dot{x}(t) = - \partial_x V(x; \alpha(t)) + \sqrt{2D}\,\eta(t)
\end{equation}
with a confining potential $V(x; \alpha) = \alpha x^2/2$ whose stiffness $\alpha(t)$ is governed by the following mean-reverting process
\begin{equation}
\dot{\alpha}(t) = - \partial_\alpha {\cal V}(\alpha) + \sqrt{2D_{\alpha}}\,\xi(t)
\end{equation}
where $\eta(t)$ and $\xi(t)$ are two zero mean, unit variance Gaussian white noises. Here, we thus consider the special case where the {\it stiffness} confining potential is defined as 
\begin{equation}
{\cal V}(\alpha) = \frac{1}{2} \mu (\alpha - \alpha_0)^2
\end{equation}
where $\alpha_0 > 0$ is required for the steady-state to be well-defined. 

As before, $P^{\rm tot}(\alpha,t) = \int dx \,P(x,\alpha,t)$ denotes the marginal probability density for the potential having a particular stiffness $\alpha$ independently of the position $x$ of the trapped particle. Starting from Eq.~\eqref{eq:en_flow_cont}, the entropy flow for this model can be written as
\begin{equation}
	\dot{S}_e(t) = \langle \alpha \rangle - \int d\alpha  \frac{\alpha^2}{D} \Xi(\alpha,t) + \frac{1}{D_\alpha} \int d\alpha \ \mathcal{J}^{\rm tot}(\alpha,t) \mathcal{V}'(\alpha)
	\label{eq:EPforOUOU}
\end{equation}
where $\langle \alpha \rangle = \int d\alpha \,\alpha P^{\rm tot}(\alpha,t)$ denotes the average potential stiffness, $\mathcal{J}^{\rm tot}$ is the probability current in \textit{stiffness space} and the marginal variances are defined as $\Xi(\alpha,t) =  \int dx \ x^2 P(x,\alpha,t)$. 

As noted before, in this continuum limit, the marginal probability density is governed by the following Fokker-Planck equation
\begin{equation}
	\partial_t P^{\rm tot}(\alpha,t) = \mathcal{L} P^{\rm tot}(\alpha,t)
\end{equation}
with $\mathcal{L}$ the linear Fokker-Planck operator \cite{Risken1996}. Therefore, the steady-state marginal probability density $P^{\rm tot}(\alpha)$ is Gaussian and is given by
\begin{equation}
	P^{\rm tot}(\alpha) = \sqrt{\frac{\mu}{2 \pi D_\alpha}} {\rm exp}\left( - \frac{\mu(\alpha-\alpha_0)^2}{2 D_\alpha} \right) \label{eq:norm_alpha}
\end{equation}
with $\mathcal{J}^{\rm tot}(\alpha)=0$ since this is an equilibrium process and the average stiffness reduces to $\langle \alpha \rangle = \alpha_0$. We are left to calculate the second term in Eq.\,(\ref{eq:EPforOUOU}) to finally obtain the entropy production. Note that the marginal variances satisfy at steady-state the following linear equation
\begin{equation}
	(\mathcal{L} - 2\alpha) \Xi(\alpha) + 2D P^{\rm tot}(\alpha) = 0 ~.
\end{equation}
Indeed, the Fokker-Planck equation for $P(x,\alpha,t)$ is written as
\begin{equation}
	\partial_t P = D\partial_x^2P + \alpha \partial_x\big[xP\big] + D_\alpha \partial_\alpha^2P + \mu\partial_\alpha\big[(\alpha-\alpha_0)P\big]
	\label{eq:FPOUOU}
\end{equation}
where we have dropped the functional dependencies for the sake of simplicity. From here, we can follow a similar procedure to that used to obtain Eq.\,\eqref{eq:G2OUMS} and find that the marginal variance $\Xi(\alpha,t)$ is governed by the following kinetic equation
\begin{equation}
\partial_t \Xi(\alpha,t) = 2DP^{\rm tot}(\alpha,t)  - 2\alpha \Xi(\alpha,t)+\partial_\alpha\big[D_\alpha\partial_\alpha\Xi(\alpha,t) + \mu(\alpha-\alpha_0)\Xi(\alpha,t)\big].
\label{eq:TDVarA}
\end{equation}
Integrating this last equation at steady-state with respect to $\alpha$ leads to $\langle\alpha x^2\rangle = D$. This itself is a remarkable result, indicating that the effect on positional fluctuations, as captured by the variance $\langle x^2 \rangle$, associated with changes in $\alpha_0$, $D_\alpha$ or $\mu$ is exactly cancelled when the displacement is rescaled by the fluctuating stiffness $\alpha(t)$, such that the scaled variance $\langle \alpha x^2 \rangle$ is independent of the stiffness dynamics.

\begin{figure}
	\centering
	\includegraphics[width=\textwidth]{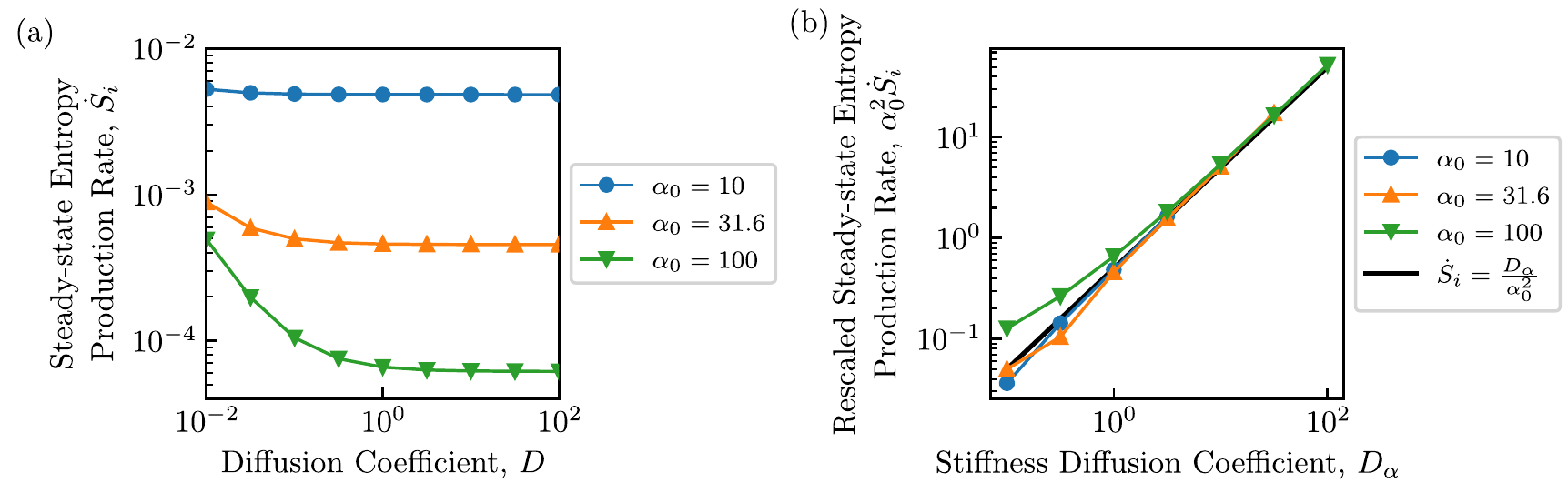}
	\caption{{\it Steady-state entropy production for a Brownian particle in a harmonic potential with continuously varying stiffness ---} (a) Steady-state entropy production rate as a function of the particle self-diffusivity $D$, for $\mu = D_{\alpha} = 1$ and different values of $\alpha_0$. The entropy production rate becomes independent of the positional diffusion coefficient for large enough values of $D$ and remains finite and non-negative at low values of $D$. (b) Steady-state entropy production rate increases with $D_\alpha$. We find that $\dot{S}_i \sim D_{\alpha}/\alpha_0^2$. Here, we set $D=\mu=1$ and vary $D_\alpha$ and $\alpha_0$.} 
	\label{fig:cont_stiff}
\end{figure}

We then multiply (\ref{eq:TDVarA}) by $\alpha$ before again integrating over $\alpha$ to obtain
\begin{equation}
	\int d\alpha \bigg[\frac{\alpha^2}{D}\Xi(\alpha)\bigg] = \langle \alpha \rangle +  \frac{1}{2D}\int d\alpha \bigg[ D_\alpha\partial_\alpha\Xi(\alpha) + \alpha\partial_\alpha\big[\mu(\alpha-\alpha_0)\Xi(\alpha)\big]\bigg] 
	\label{eq:int_step}
\end{equation}
with $\langle \alpha \rangle = \alpha_0$. Finally, we argue that the second term on the right-hand side of \eqref{eq:int_step} equation can be written as
\begin{equation}
\frac{1}{2D}\int d\alpha \bigg[ D_\alpha\partial_\alpha\Xi(\alpha) + \alpha\partial_\alpha\big[\mu(\alpha-\alpha_0)\Xi(\alpha)\big]\bigg]  =  -\frac{\mu}{2D}\big[\langle\alpha x^2\rangle - \alpha_0\langle x^2\rangle\big],
\end{equation}
noticing that the term proportional to $D_\alpha$ vanishes by imposing a sufficiently fast decay of $\partial_x P$ at $x \to \pm \infty$. Using $\langle \alpha x^2 \rangle = D$, we conclude that the entropy production rate at steady-state can be expressed as
\begin{equation}
	\lim_{t\rightarrow\infty}\dot{S}_i = -\frac{\mu}{2D}\langle (\alpha-\alpha_0)x^2\rangle = \frac{\mu\alpha_0}{2D}\bigg(\langle x^2\rangle - \frac{D}{\alpha_0}\bigg),
	\label{eq:SSEPOUOU}
\end{equation}
which is the simplest exact form for the entropy production that we can obtain here and the main result of this section. Note that the limit $D_\alpha \rightarrow 0$ represents an equilibrium limit for the system, we argue that in this case the variance of the particle position is given by $\langle x^2\rangle=D/\alpha_0$ and thus one would observe no entropy production, as expected for an equilibrium process. We have thus expressed the entropy production in this system through the difference between the particle positional variances in the fully nonequilibrium process and its equilibrium limit. A closed-form solution for the entropy production in this system relies on our ability to calculate the variance of the particle position; while this can easily be achieved numerically (see Fig.\,\ref{fig:cont_stiff}), it is not possible to write an analytical expression for it in general.

As shown in Fig.\,\ref{fig:cont_stiff}, we observe that the steady-state entropy production rate decays with increasing diffusion coefficient $D$. For low values of $D$, while Brownian motion becomes progressively weaker, fluctuations in the particle position (as captured by $\langle x^2 \rangle$) remain significant due to the existence of periods of transiently negative potential stiffnesses. As a consequence, we expect the bracketed terms in Eq.~\eqref{eq:SSEPOUOU} to remain finite as $D$ approaches $0$, leading to the observed increase of the entropy production in this limit. On the other hand, the steady-state entropy production rate converges to a finite value and becomes independent of $D$ at large enough diffusivities. When $\alpha_0,D \gg \mu, D_{\alpha}$, we effectively obtain a separation of timescales between the dynamics in $x$-space and $\alpha$-space. Assuming $\alpha_0^2 \gg D_\alpha/\mu$, the variance of the particle position is then well-approximated by the average over positive $\alpha$ of the variance of particle in a fixed potential with stiffness $\alpha$, $\langle x^2 \rangle_{\alpha} = D/\alpha$, weighted by the probability to observe such a potential stiffness $P(\alpha,t)$. Altogether, we thus expect the term in the brackets in Eq.\,\eqref{eq:SSEPOUOU} to scale like $D$ and the $D$ dependence to finally scale out of the steady-state entropy production rate. Finally, we confirm our intuition that the entropy production rate should increase with increasing values of diffusivity in $\alpha$-space and show that $\dot{S}_i \sim D_{\alpha}^{\beta}$, with $\beta \approx 1$.

Furthermore, we can verify that entropy production \eqref{eq:SSEPOUOU} is non-negative by considering
\begin{equation}
	\langle (\alpha-\alpha_0)^2 x^2 \rangle = \langle \alpha^2x^2 \rangle - 2 \alpha_0\langle \alpha x^2\rangle + \alpha_0^2\langle x^2\rangle \geq 0~,
\end{equation}
where the equality is only saturated in the deterministic limit, i.e.\ for $D =D_\alpha = 0$. Substituting once again $\langle\alpha x^2\rangle = D$ in the above equation, and using 
\begin{equation}
	\langle\alpha^2 x^2\rangle = D\alpha_0 - \frac{D\mu}{2} + \frac{\mu\alpha_0}{2}\langle x^2\rangle,
	\label{eq:a2x2}
\end{equation}
which is obtained from the Fokker-Planck equation \eqref{eq:FPOUOU}, we eventually find 
\begin{equation}
	\langle (\alpha-\alpha_0)^2 x^2 \rangle = \bigg(\alpha_0+\frac{\mu}{2}\bigg)\big(\alpha_0\langle x^2\rangle - D\big)
	\label{eq:mean5}
\end{equation}
and hence $\alpha_0 \langle x^2 \rangle \geq D$, as required.

\subsection{Fast stiffness dynamics limit}\label{sec:faststiffness}

Finding an analytical expression for the entropy production rate of a diffusive particle in a harmonic potential whose stiffness is governed by an Ornstein-Uhlenbeck process relies on our capacity to write down the variance of the particle position. To make some headway along this line, we consider the regime where the stiffness dynamics are much faster than the positional dynamics of the particle. Here, we work perturbatively and introduce a small parameter, $\varepsilon \ll 1$, characterizing the separation in timescales between the two processes, as is common practice in the literature for fast-slow dynamical systems \cite{Pavliotis2008}. 

At the level of the coupled Langevin equations, this re-scaling is written as
\begin{subequations}
\begin{align}
	\dot{x}(t) &= -\alpha x(t) + \sqrt{2D}\:\eta_x(t)\\
	\varepsilon \dot{\alpha}(t) & = -\tilde{\mu}(\alpha(t)-\alpha_0) + \sqrt{2\tilde{D}_\alpha\varepsilon}\:\eta_\alpha(t)
\end{align}
\end{subequations}
where we have taken care to re-scale the noise appropriately under the separation of timescales. The Fokker-Planck equation for the joint probability density now reads
\begin{align}
	\partial_t P(x,\alpha,t) = D\partial_x^2P(x,\alpha,t) + \alpha &\partial_x\big[xP(x,\alpha,t)\big] \nonumber \\
										    &+ \frac{\tilde{D}_\alpha}{\varepsilon} \partial_\alpha^2P(x,\alpha,t) + \frac{\tilde{\mu}}{\varepsilon}\partial_\alpha\big[(\alpha-\alpha_0)P(x,\alpha,t)\big]~,
	\label{eq:FPOUOUe}
\end{align}
corresponding to the rescaling $D_\alpha \to \tilde{D}_\alpha/\varepsilon$, $\mu \to \tilde{\mu}/\varepsilon$, which preserves the variance of the stiffness. 

In the limit $\varepsilon\rightarrow0$, it is known that $P(x,\alpha)\rightarrow P(x;\alpha_0)P^{\rm tot}(\alpha) $ where $P(x;\alpha_0)$ is the stationary distribution in the case where $\alpha\equiv\alpha_0$ and $P^{\rm tot}(\alpha) $ is the stationary marginal distribution for $\alpha$ as given in Eq.\,\eqref{eq:norm_alpha} \cite{Pavliotis2008}. For small but finite $\varepsilon$, it is useful to write the stationary probability distribution perturbatively around this limit, namely 
\begin{equation}
	P(x,\alpha) = P_0(x; \alpha_0)P^{\rm tot}(\alpha)  + \varepsilon P_1(x,\alpha),
	 \label{eq:eps_exp_p}
\end{equation}
where $P_1(x,\alpha)$ is some function of leading order $\mathcal{O}(\varepsilon^0)$ into which all higher order corrections have also been absorbed \cite{Pavliotis2008}. Note that $P_1(x,\alpha)$ should not be thought of as a probability distribution as it does not satisfy the normalization condition, rather
\begin{equation}
	\int dx\int d\alpha\: P_1(x,\alpha) = 0.
\end{equation}
We introduce the notation
\begin{equation}
	\langle \:\cdot\: \rangle_1 = \int dx \int d\alpha\: (\cdot)\: P_1(x, \alpha)
\end{equation}
whence 
\begin{equation}
	\langle x^2\rangle - \frac{D}{\alpha_0} = \varepsilon\langle x^2\rangle_1 
	\label{eq:mean1}
\end{equation}
which allows us to express the variance of the position in terms of the variance in the uncoupled problem where $\alpha \equiv \alpha_0$ and a contribution from the first-order term in $\varepsilon$. From Eq.\,\eqref{eq:SSEPOUOU}, it is clear that to compute the steady-state entropy production rate, we need to find an analytic expression for the quantity $\langle x^2\rangle_1$. Multiplying the Fokker-Planck equation \eqref{eq:FPOUOUe} by $x^2(\alpha-\alpha_0)^2$ and integrating with respect to both $x$ and $\alpha$ at steady-state leads after some straightforward algebra to the following moments relation
\begin{equation}
	D \langle(\alpha-\alpha_0)^2 \rangle - \langle\alpha(\alpha-\alpha_0)^2x^2 \rangle + \frac{\tilde{D}_\alpha}{\varepsilon} \langle x^2 \rangle - \frac{\tilde{\mu}}{\varepsilon} \langle x^2 (\alpha-\alpha_0)^2\rangle = 0
	\label{eq:momentsrelation}
\end{equation}
While we have already expressed the variance of the particle position in terms of our perturbative expansion \eqref{eq:eps_exp_p}, similarly, we write the other moments as 
\begin{subequations}
\begin{align}
	\langle(\alpha-\alpha_0)^2\rangle &= \varepsilon \langle(\alpha-\alpha_0)^2\rangle_1 + \frac{\tilde{D}_\alpha}{\tilde{\mu}} \label{eq:contmoment1} \\
	\langle (\alpha-\alpha_0)^2x^2\rangle &= \varepsilon \langle (\alpha-\alpha_0)^2x^2\rangle_1 + \frac{D\tilde{D}_\alpha}{\alpha_0\tilde{\mu}} \label{eq:contmoment2} \\
	\langle\alpha(\alpha-\alpha_0)^2x^2 \rangle &= \varepsilon \langle\alpha(\alpha-\alpha_0)^2x^2 \rangle_1 + \frac{D\tilde{D}_\alpha}{\tilde{\mu}} \label{eq:contmoment3}
\end{align}
\label{eq:contmoments}
\end{subequations}
Substituting \eqref{eq:mean1} and \eqref{eq:contmoments} in \eqref{eq:momentsrelation}, we obtain 
\begin{equation}
	\langle(\alpha-\alpha_0)^2x^2\rangle_1 = \frac{\tilde{D}_\alpha}{\tilde{\mu}}\langle x^2\rangle_1 - \frac{\varepsilon}{\tilde{\mu}}\langle \alpha(\alpha-\alpha_0)^2x^2\rangle_1.
	\label{eq:mean3}
\end{equation}
Furthermore, taking care to rescale $\mu \to \tilde{\mu}/\varepsilon$, Eq.\,\eqref{eq:mean5} can be rewritten as follows
\begin{equation}
	\langle (\alpha-\alpha_0)^2 x^2 \rangle = \left(\alpha_0+\frac{\tilde{\mu}}{2\varepsilon}\right)\alpha_0 \varepsilon \langle x^2 \rangle_1
	\label{eq:moment2}
\end{equation}

\begin{figure}[t]
	\centering
	\includegraphics[width=0.6\textwidth]{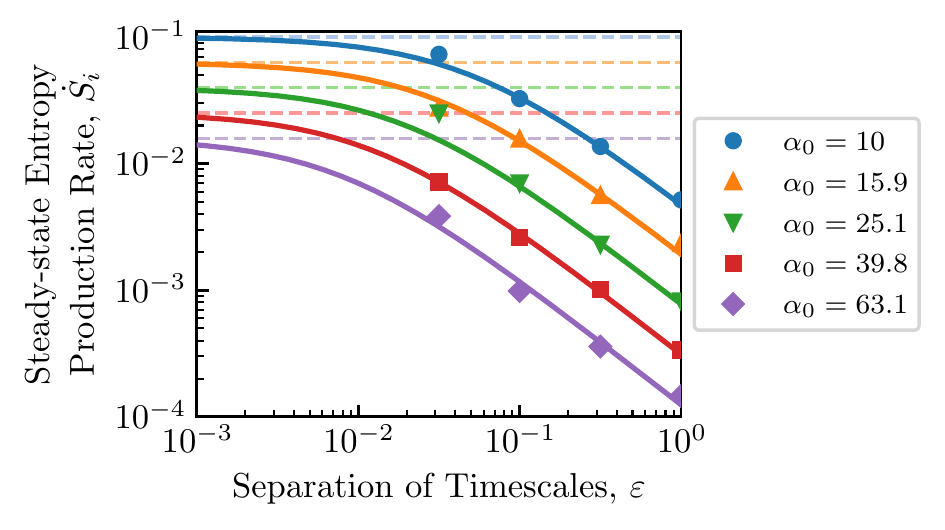}
	\caption{{\it Steady-state entropy production as a function of the separation of timescale, $\varepsilon$ ---} We fix $D=D_\alpha=\mu=1$ and $\alpha_0=10$, then vary $\varepsilon$ which represents the difference in the timescales of the two processes, introduced in Sec \ref{sec:faststiffness}. We show good agreement between numerical simulations (symbols) and analytic result \eqref{eq:SIeps} (solid lines). We also show the analytic results for the entropy production rate as $\varepsilon\rightarrow0$ (dotted lines) as given by \eqref{eq:Si_cont_eps0}.}
	\label{fig:fast_stiff}
\end{figure}

\noindent Finally, we combine \eqref{eq:contmoment2}, \eqref{eq:mean3} and \eqref{eq:moment2} to obtain a closed-form expression for $\langle x^2 \rangle_1$ valid up to order ${\cal O}(\varepsilon^2)$, 
\begin{equation}
	\langle x^2\rangle_1 = \frac{D\tilde{D}_\alpha}{\alpha_0\tilde{\mu}} \bigg[\frac{\tilde{\mu}\alpha_0}{2}+ \varepsilon\bigg(\alpha_0^2 - \frac{\tilde{D}_\alpha}{\tilde{\mu}}\bigg)+\mathcal{O}(\varepsilon^2)\bigg]^{-1}.
\end{equation} 
We conclude that the entropy production as given in Eq.~\eqref{eq:SSEPOUOU} is therefore 
\begin{equation}\label{eq:SIeps}
	\lim_{t \to \infty} \dot{S}_i = \frac{\tilde{D}_\alpha}{\alpha_0 \tilde{\mu}} \left[1 + \frac{2\varepsilon}{\alpha_0\tilde{\mu}}\bigg(\alpha_0^2 - \frac{\tilde{D}_\alpha}{\tilde{\mu}}\bigg)+\mathcal{O}(\varepsilon^2) \right]^{-1} 									   
\end{equation}
which we compare to the results of numerical simulations in Fig.\,\ref{fig:fast_stiff}. It follows that as we saw for the cases of discrete stiffness, the entropy production remains finite in the limit of fast stiffness dynamics, here $\varepsilon \to 0$. Namely, we obtain 
\begin{equation}\label{eq:Si_cont_eps0}
	\lim_{\varepsilon \to 0} \lim_{t\to\infty} \dot{S}_i(t) = \frac{\tilde{D}_\alpha}{\alpha_0 \tilde{\mu}}\,,
\end{equation}
and conclude that it scales linearly with the variance of $P^{\rm tot}(\alpha)$. 

\section{Conclusion and discussion}
\label{sec:disc}

In this work, we have established a general framework for calculating the steady-state entropy production rate of diffusive single-particle systems in time-dependent, confining potentials subject to Markovian stochastic fluctuations, including both discrete and continuous ``state spaces'' for the fluctuating potential. Our exploration has been conducted within the formalism of \cite{Gaspard2004}, reviewed in \cite{Cocconi2020}. After introducing our results for general Markovian processes, we obtain analytical results for a variety of important cases. In particular, we focus on harmonic confining potentials subject to fluctuations in the stiffness $\alpha$. 

As a first example, we study the diffusion of a particle in an intermittent harmonic potential switching {\it on} and {\it off} with a symmetric rate $k$. In this case, we conclude that the entropy production is independent of both the diffusivity $D$ of the trapped passive particle and the switching rate $k$. This remarkably simple result emerges naturally from the quadratic form of the confining potential. Indeed, one expects the steady-state positional probability density, which determines the steady-state probability current, to generically depend on both of these parameters, as we show in the simple example of a quartic potential. 

We then expanded this preliminary result to a general two-state Ornstein-Uhlenbeck process with Markovian switching. Using this model, we discussed the entropy production in a realistic model of stochastic resetting, a problem which has previously attracted attention of several groups \cite{Fuchs2016,Evans2020,Busiello2020}. Within our framework, traditional stochastic resetting is associated with infinite entropy production on the basis of a complete breakdown of time reversal symmetry \cite{Evans2020}. We reconcile this observation with the finite entropy production calculated in \cite{Fuchs2016,Busiello2020} by recognizing that the measures of dissipation used in these latter works are not directly linked to time-reversal symmetry. Thereupon, we further generalized our results on entropy production to harmonic potentials with stiffnesses controlled by an $N$-state discrete Markov process. As a direct application, we studied a simple example of a $3$-state process highlighting the emergence of a non-trivial contribution to the entropy production due to currents in the stiffness space.

Finally, we explored a model where the potential stiffness itself evolves in time according to an Ornstein-Uhlenbeck process with diffusivity $D_\alpha$ and coupling $\mu$, modeling, for instance, the diffusion of a particle confined in an optical trap whose strength is fluctuating continuously in time due to  e.g. fluctuations in laser intensity. Strikingly, we give explicit analytical results for the entropy production in the regime where the stiffness fluctuations are fast compared to the positional dynamics of the particle.

Interestingly, we observed in some cases that the entropy production remains finite upon taking limits for which the \emph{dynamics} of the trapped Brownian particle are indistinguishable from those of an equilibrium model, reminiscent of the diffusive limit for RnT particles with diverging tumbling rate \cite{Cocconi2020,GarciaMillan2021}. This phenomenon, which is sometimes referred to as an \emph{entropic anomaly} \cite{Celani2012,bo2014entropy}, is a common occurrence for systems with interacting fast and slow degrees of freedom, which points to the non-trivial correspondence between dynamic and thermodynamic features of nonequilibrium stochastic processes.

Altogether, this work forms a comprehensive study of the entropy production for single-particle systems with fluctuating potentials, which provides the foundations of a nonequilibrium thermodynamic theory of fluctuating potentials. While we have derived exact results for the case where the potential is of quadratic form, the framework developed here can readily be extended to more complex confining potentials. Further, we focus here on diffusive motion in confining fluctuating potentials but our framework itself can be generalized to a more general class of models like random acceleration processes \cite{Goldman1971,Burkhardt2000,Majumdar2010,Singh2020} or active particles including run-and-tumble particles \cite{Bertrand2018a,Bertrand2018b,GarciaMillan2021,Solon2015}, active Brownian particles \cite{Solon2015,Romanczuk2012,Bechinger2016} and active Ornstein-Uhlenbeck particles \cite{Bonilla2019,Caprini2019,Martin2021,Semeraro2021}, which will be the subject of future work. Finally, we believe that our results provide a natural framework to study the stochastic thermodynamics of colloidal systems in optical traps \cite{Grier1997,Dufresne2001,Bustamante2021}.  

\section*{Acknowledgments}
The authors thank Gunnar Pruessner for fruitful discussions. HA was supported by a Roth PhD scholarship funded by the Department of Mathematics at Imperial College London. LC acknowledges support from the Francis Crick Institute, which receives its core funding from Cancer Research UK (FC001317), the UK Medical Research Council (FC001317), and the Wellcome Trust (FC001317). 


\appendix

\section{Steady-state densities for Brownian motion in an intermittent harmonic potential}
\label{app:intermittent}

In Section \ref{sec:intermittent}, we calculate the entropy production for a particle diffusing in an intermittent harmonic potential. Exact results are known for the stationary distributions for the specific process. Therefore, our analytical result for the entropy production can be directly compared to that obtained by directly integrating Eqs.\,\eqref{eq:en_prod} and \eqref{eq:en_flow}. For completeness, we rederive here shortly the steady-state distributions following the derivations found in \cite{Zhang2017,Santra2021}. 

To do so, we start from the Fokker-Planck equations \eqref{eq:kinetic} at steady-state which read
\begin{subequations}
	\begin{align}
		0 &= D \partial^2_x P_{\rm off}(x)+ kP_{\rm on}(x) - kP_{\rm off}(x) \\
		0 &= D \partial^2_x P_{\rm on}(x) + \alpha_0 \partial_x \big[x P_{\rm on}(x)\big] + k P_{\rm off}(x) - k P_{\rm on}(x)
	\end{align}
	\label{eq:ss_fokker_planck_intermittent}%
\end{subequations}
where we have dropped the time dependence in $P_{\rm on}(x)$ and $P_{\rm off}(x)$ to denote their stationary nature. To solve these coupled equations, it is easier to work in Fourier space. Using the following convention for Fourier transforms, 
\begin{equation}
	\widehat{P}_{i} (\nu,t) =  \int_{-\infty}^{+\infty} P_i(x,t) e^{-i\nu x} dx\,,
\end{equation}
these equations read in Fourier-transformed space
\begin{subequations}
	\begin{align}
		0 &= -D \nu^2 \widehat{P}_{\rm off}(\nu)+ k\widehat{P}_{\rm on}(\nu) - k\widehat{P}_{\rm off}(\nu) \label{eq:fourier_intermittent_1}\\
		0 &= -D \nu^2 \widehat{P}_{\rm on}(\nu) - \alpha_0 \nu \partial_\nu \widehat{P}_{\rm on}(\nu) + k \widehat{P}_{\rm off}(\nu) - k \widehat{P}_{\rm on}(\nu) \label{eq:fourier_intermittent_2}
	\end{align}
	\label{eq:fourier_ss_fokker_planck_intermittent}%
\end{subequations}
Using Eq.\,\eqref{eq:fourier_intermittent_1}, we can express $\widehat{P}_{\rm off}(\nu)$ in terms of $\widehat{P}_{\rm on}(\nu)$ which allows us to write the following single differential equation for $\widehat{P}_{\rm on}(\nu)$
\begin{equation}
	\alpha_0 \partial_\nu \widehat{P}_{\rm on}(\nu) + D \nu \left[ 1 + \frac{k}{D\nu^2 + k}\right] \widehat{P}_{\rm on}(\nu) = 0~.
	\label{eq:diff_pon_intermittent}
\end{equation}
The solution to Eq.\,\eqref{eq:diff_pon_intermittent} can easily be shown to read
\begin{equation}
	\widehat{P}_{\rm on}(\nu) = \left[\frac{k}{2(k+D\nu^2)}\right]^{\frac{k}{2\alpha_0}} \exp \left[-\frac{D\nu^2}{2\alpha_0}\right]
\end{equation}
where we have used the fact that $\widehat{P}(0) =\widehat{P}_{on}(0) + \widehat{P}_{\rm off}(0) $ by conservation of probability. Finally, using Eq.\,\ref{eq:fourier_intermittent_1}, we obtain the full steady-state distribution in $\nu$-space as
\begin{equation}
	\widehat{P}(\nu) = \frac{1}{2} \left[ \frac{e^{-D\nu^2/2\alpha_0}}{(1+D\nu^2/k)^{k/2\alpha_0}} +  \frac{e^{-D\nu^2/2\alpha_0}}{(1+D\nu^2/k)^{1+k/2\alpha_0}}  \right]
\end{equation}

While it is not possible to obtain a closed-form expression for the total steady-state distribution in real space for general values of $\alpha_0$, $D$ and $k$, one can invert this relation and write $P(x)$ as the following sum of convolution integrals
\begin{equation}
	P(x) = \frac{1}{2} \left\{ \int_{-\infty}^{+\infty} dy \, f_2\left(y,\frac{k}{2\alpha_0}\right)f_1(x-y) +  \int_{-\infty}^{+\infty} dy \, f_2\left(y,1+\frac{k}{2\alpha_0}\right)f_1(x-y) \right\}  
	\label{eq:real_conv_distribution}
\end{equation}
where
\begin{subequations}
	\begin{align}
		f_1(x) &= {\cal F}^{-1}\left[ e^{-D\nu^2/2\mu_0}\right] = \frac{1}{\sqrt{2\pi D/\alpha_0}}e^{-\alpha_0 x^2/2D}\label{eq:fourier_inv_1}\\
		f_2(x, \beta) &= {\cal F}^{-1}\left[ (1+D\nu^2/k)^{\beta}\right] = \frac{\sqrt{\pi}}{\Gamma(\beta)} \left(\frac{k}{D}\frac{|x|}{2} \right)^{\beta-1/2} K_{\frac{1}{2}-\beta}\left( \sqrt{\frac{k}{D}}|x|\right) \label{eq:fourier_inv_2}
	\end{align}
	\label{eq:fourier_inverses}%
\end{subequations}
with $K_n(x)$ the modified Bessel function of the second kind.

The form of the steady-state distribution for the particle position emerges from a competition between two timescales: (i) $k^{-1}$ the timescale set by the switching rate of the intermittent confining potential and (ii) $\alpha_0^{-1}$ which sets the particle position correlation time, or equivalently, the timescale at which the particle position converges back to the center of the confining potential. As shown in Ref.\,\cite{Santra2021}, it is possible to obtain exact expressions for the steady-state distribution in some asymptotic regimes. In particular, in the limit where the switching rate is very small compared to the confining potential strength, $k\ll \alpha_0$, the steady-state distribution is given by 
\begin{equation}
	P(x) \underrel{k\ll \alpha_0}{=} \frac{1}{2} \left[ \frac{e^{-\alpha_0 x^2 /2D}}{\sqrt{2\pi D\alpha_0}}  + \frac{\sqrt{k/D}e^{k/2\alpha_0}}{4} e^{-\sqrt{k/D}|x|}\erfc \left( \frac{\sqrt{k/D}}{2 \alpha_0} - \sqrt{\frac{\alpha_0}{2D}} |x| \right) \right]
	\label{eq:ss_intermittent_asymptotic_1}
\end{equation}
leading to a central Gaussian region followed by exponential tails. Conversely, in the limit of a very fast switching rate $k\gg \alpha_0$, the steady-state distribution is given by 
\begin{equation}
	P(x) \underrel{k\gg \alpha_0}{=} \frac{e^{-\alpha_0 x^2/4D}}{\sqrt{4\pi D /\alpha_0}}
	\label{eq:ss_intermittent_asymptotic_2}
\end{equation}
which is the same as that of an equilibrium Ornstein-Uhlenbeck process with a reduced potential strength $\alpha_0/2$.

\section{Numerical analysis}
\label{app:numerical}

As shown in Eq.\,\eqref{eq:en_flow}, the entropy flow can easily be calculated if given the knowledge of the stationary distribution for the process. However, analytic forms for these stationary distributions are generically difficult to obtain. In order to confirm our analytical results, we can nonetheless resort to computing the entropy production numerically. To do so, we measure the stationary distribution (or histogram of the particle positions) for each of our models directly from simulated single particle trajectories over long times. 

In all systems, the single particle trajectories are obtained by solving the associated Langevin equation using a stochastic Runga-Kutta method with a fixed time step, $dt=10^{-5}$, for $t\in[0,10^{4}]$ \cite{Branka1999}. When considering a discrete Markov process for the fluctuating potential, $\alpha(t)$ is updated by evaluating the transition probabilities based on the switching rates and timestep. For the continuous Markov process, the stiffness itself follows a Langevin equation, which we solve using a stochastic Runga-Kutta method as above \cite{Branka1999}.

\section{Entropy production for intermittent quartic potential}
\label{app:quartic}

We consider a simple modification of the preliminary example introduced in Section \ref{sec:intermittent} in which we replace the intermittent quadratic potential by an intermittent \textit{quartic} potential, $V(x;\alpha(t))=\alpha(t) x^4/4$. The equation for the steady-state entropy production can be derived using the same procedure. The corresponding equation to (\ref{eq:SI}) is in this case
\begin{equation}
\lim_{t\rightarrow\infty}\dot{S}_i(t) = 3\alpha_{0}\int dx\,x^2 P_{\rm off}(x) = 3\alpha_{0}\Xi_{\rm off}.
\end{equation}
where $P_{\rm off}(x)$ is the steady-state joint probability density of finding an agent at position $x$ in the {\it off} state. It thus follows that the steady-state entropy production is independent of $k$ and $D$ if and only if $\Xi_{\rm off}$ is.

To show that this is not the case, suppose that $\Xi_{\rm off}$ is independent of $k$. At steady-state, we know that 
\begin{equation}\label{eq:kx4}
0 = D\partial_x^2 P_{\rm off}(x) + k P_{\rm on}(x) - k P_{\rm off}(x).
\end{equation}
Multiplying \eqref{eq:kx4} by $x^2$ and integrating over the spatial variable $x$, we find 
\begin{equation}
	 \Xi_{\rm on} = \Xi_{\rm off} - \frac{2D}{k}.
\end{equation}
If we fix $D$ and $\alpha_{0}$, then, by our earlier assumption, $\Xi_{\rm off}$ is a constant. However, this equation tells us that there exists a range for the switching rate $k$, namely $k < 2D/\Xi_{\rm off}$, for which the variance of the steady-state probability of the \textit{on} state is negative. This is a contradiction. We can use the same argument to show that $\Xi_{\rm off}$ can not be independent of $D$.  

Finally, we conclude that, in the case of an intermittent quartic potential, the entropy production must depend on both $k$ and $D$ and the independence of Eq.\,\eqref{eq:SI} {vis-\`a-vis} these two parameters is solely due to the quadratic nature of the confining potential. We argue that there is thus no reason for the entropy production to be independent of $k$ and $D$ with a more general confining potential. 


\section*{References}

\providecommand{\newblock}{}

\end{document}